\documentclass[12pt]{article}

\usepackage[utf8]{inputenc}
\usepackage{savesym}
\usepackage{cite}
\usepackage{wasysym}
\usepackage{amscd}
\usepackage{graphicx}
\usepackage{pifont}
\usepackage{float}
\usepackage[all]{xy}
\usepackage{multicol}
\usepackage{amsfonts}
\usepackage{amscd}
\usepackage{picture}
\usepackage{amssymb}
\savesymbol{iint}
\savesymbol{iiint}
\usepackage{amsmath}
\usepackage{amsthm}

\newtheorem{prop}{Proposition}
\newtheorem{lem}{Lemma}
\newtheorem{cor}{Corollary}
\restoresymbol{TXF}{iint}
\restoresymbol{TXF}{iiint}
\usepackage{color}
\usepackage{clay}
\usepackage{hyperref}
\usepackage{slashed}
\hypersetup{colorlinks=true}
\hypersetup{linkcolor=black}
\hypersetup{citecolor=black}
\hypersetup{urlcolor=black}
\usepackage{setspace}
\usepackage{multirow}
\usepackage{wrapfig}
\usepackage{verbatim}
\numberwithin{equation}{section}

\usepackage{subcaption}
\usepackage{tikz}
\usetikzlibrary{calc}
\usetikzlibrary{shapes.misc}
\tikzset{cross/.style={cross out, draw=black, minimum size=2*(#1-\pgflinewidth), inner sep=0pt, outer sep=0pt},
cross/.default={1pt}}
\usetikzlibrary{decorations.markings}
\usepackage{braket}
\usepackage{tikz-cd}
\usepackage{dsfont}
\usepackage{mathtools}

\DeclarePairedDelimiter\abs{\lvert}{\rvert}%
\DeclarePairedDelimiter\norm{\lVert}{\rVert}%

\makeatletter
\let\oldabs\abs
\def\abs{\@ifstar{\oldabs}{\oldabs*}}
\let\oldnorm\norm
\def\norm{\@ifstar{\oldnorm}{\oldnorm*}}
\makeatother

\newcommand{\Z}{\mathbb{Z}}

\newcommand{\cI}{\mathcal{I}}

\newcommand{\cO}{\mathcal{O}}
\newcommand{\cH}{\mathcal{H}}

\DeclareMathOperator{\uU}{U}
\DeclareMathOperator{\SU}{SU}

\DeclareMathOperator{\Sq}{Sq}
\DeclareMathOperator{\SO}{SO}
\DeclareMathOperator{\Sp}{Sp}

\DeclareMathOperator{\Spin}{Spin}

\DeclareMathOperator{\pt}{pt}

\DeclareMathOperator{\Hom}{Hom}

\newcommand{\tpi}{2\pi\mathrm{i}\,}
\newcommand{\RmZ}{\mathbb{R}/\mathbb{Z}}

\newtheorem*{claim}{\hypertarget{MR}{Main Result}}
\makeatletter
	\@ifdefinable{\MRtext}{\def\MRtext/{\hyperlink{MR}{main result}}}
\makeatother

\newtheorem*{remark}{Remark}
\usepackage{xcolor}
\usepackage{tablefootnote}

\usepackage{cleveref}

\theoremstyle{definition}

\begin{document}
\date{October, 2019}

\institution{Chicago}{\centerline{${}^{1}$Kadanoff Center for Theoretical Physics \& Enrico Fermi Institute, University of Chicago}}
\institution{IAS}{\centerline{${}^{2}$Institute for Advanced Study, Princeton}}
\institution{Simons}{\centerline{${}^{3}$Simons Center for Geometry and Physics, SUNY}}
\title{Anomaly Obstructions to Symmetry Preserving Gapped Phases}

\authors{Clay C\'{o}rdova\worksat{\Chicago, \IAS}\footnote{e-mail: {\tt clayc@uchicago.edu}} and
Kantaro Ohmori\worksat{\Simons,\IAS}\footnote{e-mail: {\tt komori@scgp.stonybrook.edu}}}

\abstract{Anomalies are renormalization group invariants and constrain the dynamics of quantum field theories.  We show that certain anomalies for discrete global symmetries imply that the underlying theory either spontaneously breaks its generalized global symmetry or is gapless.  We identify an obstruction, formulated in terms of the anomaly inflow action, that must vanish if a symmetry preserving gapped phase, i.e.\ a unitary topological quantum field theory, exits with the given anomaly. Our result is similar to the $2d$ Lieb-Schultz-Mattis theorem but applies more broadly to continuum theories in general spacetime dimension with various types of discrete symmetries including higher-form global symmetries.  As a particular application, we use our result to prove that certain $4d$ non-abelian gauge theories at $\theta=\pi$ cannot flow at long distances to a phase which simultaneously, preserves time-reversal symmetry, is confining, and is gapped.  We also apply our obstruction to $4d$ adjoint QCD and constrain its dynamics.}

\maketitle

\setcounter{tocdepth}{3}
\tableofcontents

\section{Introduction}

In this paper we explore the interplay of global symmetry, anomalies, and topological field theories.  Our main new results (summarized in section \ref{DiscreteSummaryIntro} below) are constraints on the discrete anomalies that can be carried by topological field theories.  This implies that quantum field theories with certain discrete anomalies do not admit symmetry preserving gapped phases.

Our results are similar to the Lieb-Schultz-Mattis (LSM) theorem \cite{Lieb:1961fr}, which implies that some $2d$ lattice systems are either gapless or have degenerate ground states under the existence of a certain symmetry.  We generalize this analysis to continuum quantum field theory in general spacetime dimensions with various types of global symmetry.  In this work we focus on discrete global symmetries and their implications, while in upcoming work \cite{WIP} we discuss examples with continuous global symmetry making contact with the examples of ``symmetry enforced gaplessness''  discussed in \cite{Wang:2014lca, Wang:2016gqj, Sodemann:2016mib, Wang:2017txt}, as well as the examples of LSM-type theorems established in \cite{Kobayashi:2018yuk}.   

\subsection{Symmetries and Topological Operators}

Global symmetry is one of the few universally applicable tools to constrain the non-perturbative dynamics of quantum field theories.   In its most elementary incarnation, global symmetry arises from a conserved current operator $J$ which is a closed $(d-1)$-form.\footnote{The ordinary vector current is the Hodge dual, i.e.\ $dJ=0 \leftrightarrow \partial^{\mu}(*J)_{\mu}=0$.  We prefer the presentation in terms of closed forms since it makes the extension to higher-form symmetry more natural.}  The current gives rise to a conserved charge, which organizes the spectrum into multiplets.

Given a system with a global symmetry it is fruitful to couple it to background gauge fields $A$  by including in the action a term 
\begin{equation}
\mathcal{S} \supset i\int A\wedge J~.
\end{equation}
The variable $A$ is a fixed classical source, and the coupling above leads to a partition function $Z[A]$. For small $A$ (i.e.\ topologically trivial connections which are non-zero in a small region) the partition function $Z[A]$ may be viewed as a generating function for the correlation functions of the current operator $J$.  By contrast, the behavior of the partition function in topologically non-trivial backgrounds provides a convenient way to encode information about the theory that cannot be easily accessed using the correlation functions of local operators.  Our analysis below will center on such topologically non-trivial background fields.  
  
\begin{figure}[t]
	\centering
	\begin{tikzpicture}[scale = .6,thick,baseline = 0]
		\shade[ball color = gray!40, opacity = 0.4] (0,0) circle (2cm);
		\draw (0,0) circle (2cm);
		\draw (-2,0) arc (180:360:2 and 0.6);
		\draw[dashed] (2,0) arc (0:180:2 and 0.6);
		\fill[fill=black] (0,0) circle (3pt);
		\node[anchor = west] at (0,0) {$\mathcal{O}$};
		\node at (-2cm,2cm) {$U_g$};
		\fill[fill=black] (3cm,0) circle (0pt);
	\end{tikzpicture}
	=
	\begin{tikzpicture}[scale = .6,thick,baseline = 0]
		\fill[fill=black] (0,0) circle (3pt);
		\fill[fill=black] (-1cm,0) circle (0pt);
		\node[anchor = west] at (0,0) {$g(\mathcal{O})$};
	\end{tikzpicture}
	\caption{The action of a $0$-form symmetry operator $U_g$ associated to the group element $g\in G^{(0)}$ on a local operator $\mathcal{O}$. When $U_g$ surrounds $\mathcal{O}$ it converts $\mathcal{O}$ to $g(\mathcal{O})$.}
	\label{fig:0-form_action}
\end{figure}
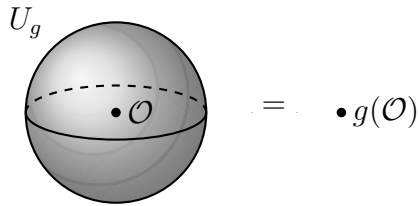

This basic paradigm of global symmetry can be generalized in several directions.  One possibility is to consider discrete global symmetries.  These are not related to conserved currents and hence we require a more abstract point of view to define the symmetry.  One way to proceed is to use symmetry operators following for instance the discussion in \cite{Gaiotto:2014kfa} and references therein.  Given a symmetry group $G$ and an element $g\in G$ there is an extended operator, a symmetry defect, $U_{g}$ of codimension one.  This operator has two key features:
\begin{itemize}
\item It is topological:  In a $d$-dimensional QFT, the operator $U_{g}$ can be defined on any manifold $\Sigma_{d-1}$ of dimension $d-1$.   The correlation functions of the $U_{g}$ operators depend only topologically on the manifold $\Sigma_{d-1}.$ This means that the symmetry defect can be deformed arbitrarily in spacetime provided it does not cross other operators.
\item It implements the $g$ action:  If the defect $U_{g}$ is deformed to cross a local operator $\mathcal{O}$, it modifies that operator to $g(\mathcal{O})$.  See Figure \ref{fig:0-form_action}.
\end{itemize}
In the special case of continuous global symmetries, the topological charge operators above are given by exponentiated integrals of the current.  However in the context of general symmetries, it is the topological operators themselves that define the symmetry of a QFT.  

Correlation functions of symmetry defects enable us to couple systems with discrete global symmetry to background gauge fields.\footnote{Mathematically, a gauge field $A$ for a discrete gauge group $G$ on the spacetime manifold $M$ can be viewed as a cocycle $A\in C^{1}(M,G).$  This cocycle is Poincar\'{e} dual to the network of symmetry defects. The gauge equivalence class is given by the cohomology class $[A]\in H^1(M,G)$.}  Note that unlike the case of continuous symmetry, there is no notion of small background fields: any background that is gauge-inequivalent to the trivial background is topologically non-trivial.  

\begin{figure}[t]
	\centering
	\begin{tikzpicture}[scale = .6,thick,baseline = 0]
		\draw (0,0) coordinate (0) arc (-180:180:2 and 1);
		\node (U) at ($(0)+(-.5,.5)$) {$U_g$};
		\coordinate (c) at ($(0)+(2,0)$);
		\draw[line width=5pt,white] (0) ++ (1,-.7) .. controls ($(c)+(-.5,0)$) and ($(c)+(1,1)$) .. ++ (3,4);
		\draw (0) ++ (1,-.7) coordinate (1) .. controls ($(c)+(-.5,0)$) and ($(c)+(1,1)$) .. ++ (3,4) coordinate(2);
		\draw (1) ++ (-.2,-.25) -- ++(-.8,-1);
		\node (L) at ($(2)+(-1,-.2)$) {$\mathcal{O}$};
	\end{tikzpicture}
	$= P(\mathcal{O}, g)\times $
	\begin{tikzpicture}[scale = .6,thick, baseline = 0]
		\draw (0,0) coordinate (0);
		\coordinate (c) at ($(0)+(2,0)$);
		\draw (0) ++ (1,-.7) coordinate (1) .. controls ($(c)+(-.5,0)$) and ($(c)+(1,1)$) .. ++ (3,4) coordinate(2);
		\draw (1) -- ++(-.2,-.25) -- ++(-.8,-1);
		\node (L) at ($(2)+(-1,-.2)$) {$\mathcal{O}$};
	\end{tikzpicture}
	\caption{The action of a $p$-form ($p\ge1$) symmetry operator $U_g$ associated to the group element $g\in G^{(p)}$ on a $p$-dimensional extended operator $\mathcal{O}$. When $U_g$ surrounds $L_p$, it acts by a phase $P(\mathcal{O},g)$ satisfying the group law $P(\mathcal{O},g_1)P(\mathcal{O},g_2) = P(\mathcal{O},g_1g_2)$. The extended operator $\mathcal{O}$ is charged under the symmetry group element $g$ when $P(\mathcal{O},g)\neq 1$. }
	\label{fig:p-form_action}
\end{figure}
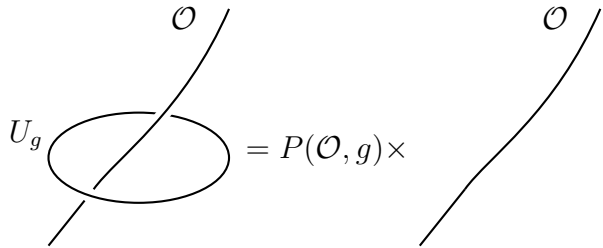

Another generalization which we will use below are so-called higher-form global symmetries \cite{Gaiotto:2014kfa}.  These may be similarly defined by the existence of topological operators of higher codimension.  Thus, a $p$-form global symmetry group $G^{(p)}$ is defined by the existence of topological operators of codimension $p+1$ for each element $g\in G^{(p)}$.  These operators also obey a fusion algebra associated to the group law
\begin{equation}
U_{g_{1}}U_{g_{2}}=U_{g_{1}g_{2}}~.
\end{equation}
The special case $p=0$ are the ordinary global symmetries defined above.  This is the only case where non-abelian groups are encountered.  The $p$-form global symmetry for $p>0$ is always an Abelian group.  For general $p$, the operators in the theory that are charged under the symmetry are extended operators of dimension $p$.  (See Figure \ref{fig:p-form_action}).

The special case of 1-form symmetry is frequently encountered, and we list several pertinent examples below.
\begin{itemize}

\item $4d$ Free Maxwell theory has 1-form global symmetry $U(1)^{(1)}\times U(1)^{(1)}$ that measures the electric and magnetic charges of Wilson-'t Hooft lines.

\item Non-Abelian gauge theories with simple gauge group $H$ and matter uncharged under the center of the gauge group $Z(H)$ have a discrete 1-form global symmetry $Z(H)^{(1)}$ measuring the charge of Wilson lines.  The behavior of this 1-form symmetry at long-distances is the order parameter for confinement \cite{Gaiotto:2014kfa}: a confined phase is symmetry preserving, a deconfined phase spontaneously breaks this symmetry.\footnote{To make the connection more precise, the area law for Wilson loops in a confining phase implies that a fundamental Wilson line which is infinite in extent has vanishing expectation value.  On the other hand a perimeter law can be viewed as a counterterm along the Wilson, so in this case infinitely long Wilson lines have non-zero expectation value.  Since infinite Wilson lines are charged under the 1-form symmetry, this establishes the connection between confinement and 1-form symmetry order parameters.}  

\item $3d$ Topological field theories have a 1-form global symmetry with charge operators given by the subset of Abelian anyons in the theory \cite{Gaiotto:2014kfa}. Many aspects of the global symmetry are encoded in the basic braiding and fusion data of the topological field theory see e.g.\ \cite{Barkeshli:2014cna, Benini:2018reh, Hsin:2018vcg} for a detailed discussion.

\end{itemize}

Again it is fruitful to couple a theory with $p$-form global symmetry to a background gauge field.  For a $p$-form global symmetry the appropriate background is a $p+1$-form gauge field.  More abstractly, for a QFT on the spacetime manifold $M$ and discrete higher-form symmetry $G^{(p)}$, gauge equivalencies classes of higher form fields are given by cohomology classes $H^{p+1}(M,G^{(p)}).$  

The collection of $p$-form symmetries of a quantum field theory may form a rich algebraic structure, where for instance fusions of $p$-form symmetry defects yield other $q$-form symmetry defects with $q>p$.  In this case the symmetries naturally form a higher-group structure as discussed in detail in \cite{Tachikawa:2017gyf, Cordova:2018cvg, Benini:2018reh, Hsin:2019fhf}.  We will not need the detailed structure of the background fields appropriate for these more general concepts of symmetry, but our analysis below will also apply to models with these embellishments.   

Finally, a relativistic quantum field theory always has Poincar\'{e} symmetry and hence may be coupled to a background metric on the spacetime manifold $M$.  Depending on the details of the model, the spacetime symmetry may require discrete refinements, which will be significant in our examples.  Several prominent cases are theories with fermions require which a choice of spin structure on $M$, and theories with time-reversal symmetry $\mathsf{T}$, which can be formulated on unorientable spacetime manifolds $M$.\footnote{In theories with both fermions and $\mathsf{T}$ one encounters $Pin^{\pm}$ structures on $M$ with $\mathsf{T}^{2}=(-1)^{F}$ yielding $Pin^{+}$ and $\mathsf{T}^{2}=1$ yielding $Pin^{-}$.}  (In certain cases, there can also be an interplay between the algebra of spacetime and internal symmetries.  See e.g.\ \cite{Cordova:2017kue}.)

In our discussion below, we will schematically denote the collection of all possible background fields for global and spacetime symmetries (including possible discrete geometric data) by $A$. Our main technical analysis will be to investigate some general properties of the partition function $Z[A]$ of a topological field theory coupled to the most general background fields.

\subsection{'t Hooft Anomalies of Global Symmetries}\label{sec:'thooftrev}

A subtle aspect of symmetry in quantum field theory is that it may have 't Hooft anomalies. We describe the essential technical framework following the recent discussion in \cite{Cordova:2019jnf, Cordova:2019uob}.  

Anomalies may be described as mild violations of gauge invariance for the background fields $A$ entering the partition function $Z[A]$.  Let us denote the gauge transformation of the background field $A$ as $A\rightarrow A^{\lambda}$ where $\lambda$ denotes the collection of all possible gauge parameters.  We say that a theory has an 't Hooft anomaly if the partition function $Z[A]$ transforms with a non-trivial phase under gauge transformations
\begin{equation}
Z[A^{\lambda}]=Z[A]\exp\left(-2\pi i \int _{M} \alpha (\lambda, A)\right)~.
\end{equation}
Here, $\alpha(\lambda, A)$ is a local functional of the gauge fields $A$ and gauge parameter $\lambda$.  

The phase $\alpha(\lambda, A)$ above is only significant if it cannot be removed by adjusting the definition of the partition function $Z[A]$ by local counterterms. Thus, the non-trivial data in the 't Hooft anomaly is a kind of cohomology class defined by the phase $\alpha(\lambda, A)$ modulo such redefinitions by counterterms.

An essential observation about anomalies is that they are rigid, i.e.\ the appropriate cohomology group is discrete.  This means that the anomaly of a quantum field theory is preserved under any continuous deformation of the parameters including in particular renormalization group flow.  This is the content of 't Hooft anomaly matching \cite{tHooft:1979rat}.  If a theory at short distances has a non-trivial anomaly, then at long-distances the same anomaly must be reproduced.  In particular, a theory with an 't Hooft anomaly cannot flow at long distances to a trivially gapped theory.\footnote{A trivially gapped theory has an energy gap, and has the feature that it has a unique ground state on any finite volume spatial manifold.  This means that it does not include any topological degrees of freedom.} 

It is useful to have a general characterization of possible anomalies of $d$-dimensional QFT with a given global symmetry.  This is provided by anomaly inflow \cite{Callan:1984sa}.  In a modern presentation, the anomalous phase $\alpha(\lambda,A)$ arises from a classical gauge-invariant Lagrangian $\omega(A)$ in $d+1$ spacetime dimensions.  The $d$-dimensional physical spacetime manifold $M$ is viewed as the boundary of a $d+1$ manifold $N$ and all background fields $A$ are extended from $M$ to $N$.  We have
\begin{equation}\label{omegadef}
\exp\left(2\pi i \int_{N}\omega(A^{\lambda})-2\pi i \int_{N}\omega(A)\right)=\exp\left(2\pi i \int_{M}\alpha(\lambda, A)\right)~.
\end{equation}
Thus, the anomaly is characterized by a local action of classical background fields but in $d+1$ spacetime dimensions.\footnote{Such local classical actions are sometimes called \emph{invertible field theories} in the terminology introduced in \cite{Freed:2004yc}. Here,``locality" does not necessarily mean that we have a convenient and explicit expression of $\omega$, but rather that this action obeys certain cutting and gluing rules. In condensed matter physics, the low energy limit of an SPT is also described by such a theory.  We use these terms synonymously below.}  In particular, on closed $(d+1)$-manifolds the anomaly theory yields a well-defined partition function. Our new results below will involving constraining properties of topological field theories by directly evaluating the partition function of the anomaly theory.

Working in the paradigm of anomaly inflow thus means that to classify possible anomalies of $d$-dimensional QFTs with a given global symmetry, we must determine all local classical actions in $d+1$ spacetime dimensions depending on the background fields $A$.  Partition functions satisfying these criteria as well as precise versions of unitarity and locality have been studied \cite{KTalk1, Kapustin:2014tfa, Kapustin:2014gma, Kapustin:2014dxa, KTalk, Freed:2016rqq, Gaiotto:2017zba, Yonekura:2018ufj}.  They are elements of certain dual cobordism groups meaning that they are phase-valued functions from certain equivalence classes of manifolds equipped with gauge fields.\footnote{The type of $(d+1)$-manifold considered in the cobordism theory depends on the discrete details of the spacetime symmetry e.g. the existence of $\mathsf{T}$ or $(-1)^{F}$.}   We will make use of some aspects of this classification in our general discussion below. This classification leads broadly to two types of anomalies:
\begin{itemize}
\item Infinite order anomalies:  These are anomalies with an action labelled by an integer.  They are the most familiar and well-known examples of anomalies and often are first encountered in 1-loop diagrams in even-dimensional field theories (see e.g.\ \cite{AlvarezGaume:1984dr} and references therein).  The inflow action $\omega(A)$ for these type of anomalies is always an appropriate Chern-Simons term, where the integer plays the role of the level.  Anomalies of this type only arise for continuous symmetry groups.

\item Finite order anomalies:  These more subtle anomalies have an action with an overall coefficient that takes a finite set of possible distinct values. They may arise for discrete or continuous symmetry groups.  Some well-known examples are the parity anomaly in $3d$ field theories \cite{Niemi:1983rq,Redlich:1983kn,Redlich:1983dv,vanderBij:1986vn,AlvarezGaume:1984nf}, but many others have been recently discussed and applied to constrain the phases of gauge theories.  In mathematical terms, a discrete anomaly is a torsion class of a $d+1$ dimensional cobordism group.  In this paper we focus on aspects of these anomalies for the particular case of discrete symmetry groups.  The case of continuous symmetry groups will be discussed in \cite{WIP}.
\end{itemize}

\subsection{Phases of Quantum Field Theories}

Since anomalies match under renormalization group flow, it is interesting to ask what kind of phases of quantum field theories can carry a given anomaly.  We distinguish several possibilities based on the properties of the vacuum:
\begin{itemize}
\item  Symmetry Preserving Gapless Phases:  There is no energy gap above the ground state, but the expectation value of all charged operators vanish. At long distances the theory is scale invariant.  

\item Symmetry Preserving Gapped Phases:  There is an energy gap above the ground state, and the expectation value of all charged operators vanish. At long distances the theory is topological.

\item Spontaneous Symmetry Breaking Phases:  The vacuum spontaneously breaks the global symmetry.   This means that charged operators have non-vanishing expectation values.  The implications depend on the form degree of the broken symmetry:
\begin{itemize}
\item In the case of a 0-form symmetry, spontaneous symmetry breaking leads to degenerate ground states in the Hilbert space on $\mathbb{R}^{d-1}$.  If the symmetry is continuous the spectrum is gapless and contains Nambu-Goldstone bosons.  If the symmetry is discrete there are still degenerate vacua, but not necessarily a gapless spectrum.
\item In the case of spontaneous symmetry breaking of a $p$-form symmetry with $p>0$, there are extended $p$-dimensional operators with non-trivial correlations at large distances.  In this case, the broken symmetry does not imply degenerate ground states on $\mathbb{R}^{d-1}$, but rather implies that there is degeneracy on spatial slices $\mathbb{R}^{d-1-p}\times T^{p},$ where $T^{p}$ is a large $p$-torus.\footnote{See also \cite{Armas:2018atq,Armas:2018zbe,Armas:2019sbe,Delacretaz:2019brr} for recent discussion of spontaneous symmetry breaking of higher-form symmetry.}
\end{itemize}
\end{itemize}
Thus, given an anomaly theory $\omega(A),$ we would like to understand which possible phases above can saturate it.  

Broadly speaking, it is believed that anomalies can always be saturated by symmetry preserving gapless phases.  Indeed, many anomalies first arise in this way in theories of massless fermions or gauge fields with chiral field strengths (see e.g.\ \cite{AlvarezGaume:1984dr}.).  In particular the anomalies of infinite order characterized by inflow of Chern-Simons terms arise this way.  

It is also believed that anomalies can always be saturated by an appropriate pattern of spontaneous symmetry breaking.  For instance, in the case of continuous ordinary global symmetries, anomalies are matched in the spontaneously broken phase by appropriate Wess-Zumino interaction terms for the Nambu-Goldstone bosons (see e.g.\ \cite{Wess:1971yu, Witten:1983tw} for foundational presentations).  Meanwhile, anomalies in the spontaneously broken phase of discrete theories may lead to different local counterterms in the vacua related by the action of the broken charges.  Often, these in turn imply that the theories on domain walls interpolating between these vacua themselves have non-trivial anomalies.   For recently studied examples in the context of $4d$ QCD and applications to $3d$ dualities see \cite{Gaiotto:2017tne, Anber:2018xek, Hsin:2018vcg, Bashmakov:2018ghn, Cordova:2019uob}.

The final possibility, symmetry preserving gapped phases, is the most constraining.   As we describe in detail, this type of phase may or may not exist depending on the detailed nature of the anomaly in question.  Our main results will be new obstructions that prevent a given anomaly for a discrete global symmetry from admitting a symmetry preserved gapped phase.  These obstructions are derived in sections  \ref{sec:symobs} and summarized in sections \ref{DiscreteSummaryIntro}.   In related work \cite{WIP} we discuss similar obstructions for continuous symmetry groups.

The fact that symmetry preserved gapped phases are highly constrained has been known implicitly from early work on anomalies.  For instance, in the most familiar case of chiral anomalies, which are infinite order and arise from one-loop diagrams in theories of free fields, one can rigorously establish that a non-zero anomaly implies that all phases of the theory are necessarily gapless by showing that such anomaly coefficients imply that the correlation functions of local current operators are non-zero at separated points (see e.g.\ \cite{Coleman:1982yg}).

Since infinite order anomalies are known to imply gaplessness, we concentrate our attention on the more subtle case of finite order anomalies.   Let us give two illustrative examples:
\begin{itemize}
\item Consider time-reversal anomalies of three-dimensional systems with fermions where $\mathsf{T}^{2}=(-1)^{F}$.   These anomalies have a $\mathbb{Z}_{16}$ classification \cite{KTalk, Fidkowski:2013jua, Chen:2013jha, Wang:2014lca, Metlitski:2014xqa, Kapustin:2014dxa, Hsieh:2015xaa}.  One can realize a given anomaly $\nu \in \mathbb{Z}_{16}$ by a gapless phase of $k$ Majorana fermions with $\nu=k$ modulo $16$ \cite{Witten:2015aba}.  However, it is also possible to obtain the same anomaly from a gapped system.  For instance, the topological Chern-Simons theory $SO(N)_{N}$ has time-reversal symmetry due to level-rank duality $SO(N)_{+N}\leftrightarrow SO(N)_{-N}$ \cite{Aharony:2016jvv}  and realizes the anomaly $\nu=N$ modulo 16 \cite{Wang:2016qkb, Cheng:2017vja}.\footnote{There are also $\mathsf{T}$-invariant TQFTs with gauge group $O(N)$ that realize any value of $\nu$ \cite{Cordova:2017vab, Cordova:2017kue}.}  Thus in this case, the anomaly does not require that symmetry preserving phases are gapless.

\item As another example, consider instead $SU(N)$ Yang-Mills theory (without matter) at the special value of the $\theta$-angle, $\theta=\pi$.  This theory has time-reversal symmetry $\mathsf{T}$, and a one-form symmetry $\mathbb{Z}_{N}^{(1)}$ measuring the charges of Wilson lines \cite{Gaiotto:2014kfa}.  In \cite{Gaiotto:2017yup}, it was demonstrated that at even $N$ these symmetries have a mixed 't Hooft anomaly.  

As is generally the case with anomaly matching, this anomaly implies that at long distances the theory cannot be trivially gapped.  Conventional lore is that at long-distances $\mathsf{T}$ is spontaneously broken leading to two vacua.  However, especially for small $N$ the physics is somewhat uncertain and more exotic phases may be contemplated.  In particular, one might wonder if the infrared might consist of a confining theory (i.e.\ unbroken 1-form symmetry) with a unique ground state and hence unbroken $\mathsf{T}$.  Using the results summarized in section \ref{DiscreteSummaryIntro}, we prove that such a phase, if dynamically realized, is necessarily gapless.

\end{itemize}

\subsection{Summary of Results and Examples}\label{DiscreteSummaryIntro}

Let us now summarize our main result derived in section  \ref{sec:symobs} below.  It applies to $d$-dimensional theories with $p$-form global symmetry group $G^{(p)}$, and is characterized abstractly in terms of the anomaly theory and its Lagrangian $\omega(A)$ introduced above.    

We will evaluate the anomaly on a $(d+1)$-manifold, of a specific type called a mapping torus.  This is constructed from a $d$-manifold $M$ and a diffeomorphism $f: M\rightarrow M$, by identifying  $[0,1]\times M$ via $(0,m)\sim (1, f(m)).$  We denote such a space as $S^{1}\times_{f}M$. (The diffeomorphism $f$ can be trivial (i.e. the identity map) when the anomaly does not involve the spacetime symmetry.)\footnote{Mapping tori are common in the study of anomalies since, when the infinite order anomalies are absent, the anomalous phase $\alpha(\lambda, A)$ can be detected on such a manifold.  See the discussion in section \ref{sec:MappingTorus}.} We have the following general result:

\begin{claim}\label{obs1}
	Consider a $(d+1)$-manifold $N^{d+1}$ which is a mapping torus $S^1\times_f(S^{p_1+1}\times S^{p_2+1})$ with $p_{1}+p_{2}=d-2$.\footnote{We assume that if any of the $p_{i}=0$, we choose the bounding (NS) spin structure for the corresponding factor, when fermionic theories are considered.  However, the spin structure along the first $S^1$ direction can be either bounding or non-bounding (R).}
We choose any elements of the internal (non-spacetime) global symmetry from the subgroups that are not spontaneously broken:
\begin{equation}
g_0 \in G^{(0)}~,\hspace{.1in} g_1 \in G^{(p_1)}~,\hspace{.1in} g_2 \in G^{(p_2)}~,\hspace{.1in} g_3 \in G^{(p_1+1)}~,\hspace{.1in} g_4 \in G^{(p_2+1)}~.
\end{equation}
(The $g_{i}$ above need not be distinct.)  Correspondingly on $N^{d+1}$ we activate background fields $A = (A_1,A_2,B_0,B_1,B_2)$ defined by:
\begin{equation}
A_1 = g_1[S^{p_1+1}]~,\hspace{.1in} A_2 = g_2[S^{p_2+1}]~,\hspace{.1in} B_0 = g_0[S^1]~,\hspace{.1in} B_1 = g_3 [S^1\times S^{p_1+1}]~,\hspace{.1in} B_2 = g_4 [S^1\times S^{p_2+1}]~.
\end{equation}
Here, the notation is that if $g\in G^{(p)}$ and $X^{p+1}$ is an $p+1$-manifold, then $g [X^{p+1}] \in H^{p+1}(N^{d+1}, G^{(p)})$ indicates the corresponding cohomology class.\footnote{For these backgrounds to make sense on $N^{d+1}$, the cohomology classes $g_1[S^{p_1+1}]$, $g_2[S^{p_2+1}]$, $g_3[S^{p_1+1}]$ and $g_4[S^{p_2+1}]$ on $S^{p_1+1}\times S^{p_2+1}$ need to be invariant under $f$. When $p_1=p_2$, only the invariance of $g_1[S^{p_1+1}]+g_2[S^{p_2+1}]$ and $g_3[S^{p_1+1}]+g_4[S^{p_2+1}]$ is necessary.} If $g_{i} \neq 1 \in G^{(p_{i})}$ for $i=1,2$ above, we further assume that $p_{i}\leq 1~.$ 

 Then, the anomaly of any unitary topological quantum field theory satisfies:
\begin{equation}\label{obstruction}
\exp\left(2\pi i \int_{N^{d+1}} \omega(A)\right)=1~.
\end{equation}
In particular, if an anomaly $\omega$ does not obey \eqref{obstruction}, then no TQFT with anomaly $\omega$ exists.
\end{claim}

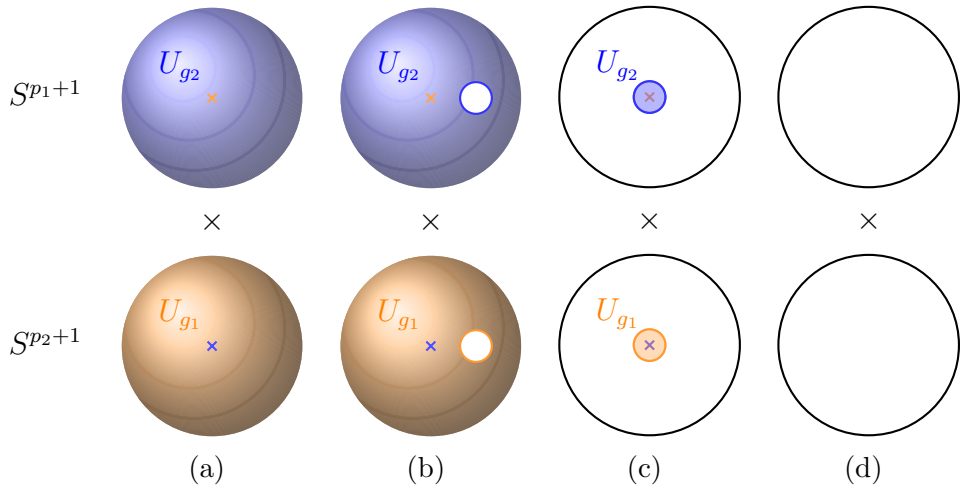
\begin{figure}[t]
	\centering
		\begin{tikzpicture}[scale = .6,thick,baseline = -130]
			\node at ($(0,0)+(-3cm,0)$) {$S^{p_1+1}$};
			\node at ($(0,-5.5cm)+(-3cm,0)$) {$S^{p_2+1}$};
		\end{tikzpicture}
	\subcaptionbox{}[.17\linewidth]{
		\begin{tikzpicture}[scale = .6,thick,baseline = 0]
			\shade[ball color = blue!40, opacity = 0.4] (0,0) coordinate(0) circle (2cm);
			\node[cross = 2.5pt,orange!70] at (0) {};
			\node[color=blue] at ($(0)+(-.7cm,.7cm)$) {$U_{g_2}$};
			\shade[ball color = orange!50, opacity = 0.5] (0,-5.5cm) coordinate(1) circle (2cm);
			\node[cross = 2.5pt,blue!70] at (1) {};
			\node[color=orange] at ($(1)+(-.7cm,.7cm)$) {$U_{g_1}$};
			\node at ($(0)!.5!(1)$) {$\times$};
		\end{tikzpicture}
	}
	\subcaptionbox{}[.17\linewidth]{
		\begin{tikzpicture}[scale = .6,thick,baseline = 0]
			\shade[ball color = blue!40, opacity = 0.4] (0,0) coordinate(0) circle (2cm);
			\node[cross = 2.5pt,orange!70] at (0) {};
			\node[color=blue] at ($(0)+(-.7cm,.7cm)$) {$U_{g_2}$};
			\draw[fill=white,draw=blue!80] ($(0)+(1cm,0)$) circle(10pt);
			\shade[ball color = orange!50, opacity = 0.5] (0,-5.5cm) coordinate(1) circle (2cm);
			\node[cross = 2.5pt,blue!70] at (1) {};
			\node[color=orange] at ($(1)+(-.7cm,.7cm)$) {$U_{g_1}$};
			\draw[fill=white,draw=orange!80] ($(1)+(1cm,0)$) circle(10pt);
			\node at ($(0)!.5!(1)$) {$\times$};
	\end{tikzpicture}
	}
	\subcaptionbox{}[.17\linewidth]{
		\begin{tikzpicture}[scale = .6,thick,baseline = 0]
			\draw[] (0,0) coordinate(0) circle (2cm);
			\node[cross = 2.5pt,orange!70] at (0) {};
			\node[color=blue] at ($(0)+(-.7cm,.7cm)$) {$U_{g_2}$};
			\draw[fill=blue!70,opacity=0.4] (0) circle(10pt);
			\draw[draw=blue!80] (0) circle(10pt);
			\draw[] (0,-5.5cm) coordinate(1) circle (2cm);
			\node[cross = 2.5pt,blue!70] at (1) {};
			\node[color=orange] at ($(1)+(-.7cm,.7cm)$) {$U_{g_1}$};
			\draw[fill=orange!70,opacity=0.4,draw=orange!80] (1) circle(10pt);
			\draw[draw=orange!80] (1) circle(10pt);
			\node at ($(0)!.5!(1)$) {$\times$};
	\end{tikzpicture}
	}
	\subcaptionbox{}[.17\linewidth]{
		\begin{tikzpicture}[scale = .6,thick,baseline = 0]
			\draw[] (0,0) coordinate(0) circle (2cm);
			\draw[] (0,-5.5cm) coordinate(1) circle (2cm);
			\node at ($(0)!.5!(1)$) {$\times$};
	\end{tikzpicture}
	}
	\caption{(a): The right hand side of \eqref{partvan}. The symmetry defect $U_{g_1}$ corresponding to the background $A_1$ is localized at a point in $S^{p_1+1}$ and wraps $S^{p_2+1}$, while the defect $U_{g_2}$ corresponding to $A_2$ is localized at a point in $S^{p_2+1}$ and wraps $S^{p_1+1}$.
	(b): Assuming that $g_1$ and $g_2$ are not spontaneously broken, the operators $U_{g_1}$ and $U_{g_2}$ admit topological boundary conditions with which the correlation function does not change.  (See section \ref{bcsec}, for details.  In particular these ``boundary conditions" need not satisfy Cardy-like constraints.)
	(c): Deforming the boundaries, the operators can be contracted to their intersection point. This contraction defines a local operator.
(d): Finally, projecting onto a sector where this operator acts as the identity, the local operator is identified with a nonzero number. Therefore \eqref{partvan} implies that the partition function on $S^{p_1+1}\times S^{p_2+1}$ vanishes, which contradicts with unitarity.}
\label{fig:partvan}
\end{figure}

We establish this statement by first noting that, if the anomaly is non-trivial on such a mapping torus it implies that the partition function of the $d$-dimensional theory $S^{p_{1}+1}\times S^{p_{2}+1}$ with symmetry backgrounds vanishes:
\begin{equation}\label{partvan}
Z[S^{p_{1}+1}\times S^{p_{2}+1}, (A_{1},A_{2})]=0~.
\end{equation}
We then prove that in a symmetry preserving TQFT, the background fields $A_{i}$ can be removed by cutting open the associated symmetry defects (See Figure \ref{fig:partvan}). Thus, \eqref{partvan} implies that the partition function without backgrounds vanishes, and this conflicts with unitarity.

By contrast, gapless or spontaneous symmetry breaking phases can easily accommodate \eqref{partvan}.  For instance, a gapless theory theory can have fermion zero modes in certain symmetry backgrounds which can lead to a vanishing partition function.  Meanwhile, if the symmetry is spontaneously broken and the symtem is gapped the resulting IR topological theory typically includes a dynamical gauge field $c$ obeying the equation of motion $\delta c\sim A$.
This means that the partition function vanishes when $A$ is non-trivial cohomology class.

In section \ref{disexsec}, we apply this analysis to several classes of examples.  In particular, we explain how our obstruction reproduces the original LSM theorem.   We also show how our obstruction applies to the 1-form symmetry in $3d$ field theories, as well as to discrete gauge theories.  Finally, we apply our results to some $4d$ gauge theories including pure Yang-Mills theory discussed above, and adjoint QCD which has recently been investigated in \cite{Shifman:2013yca, Anber:2018tcj, Cordova:2018acb, Bi:2018xvr,  Wan:2018djl, Poppitz:2019fnp}.  In particular this allows us to exclude some previously contemplated exotic phases discussed in \cite{Gaiotto:2017yup} and \cite{Bi:2018xvr}.

\subsection{Comparison to Symmetry Extension and Further Directions}\label{SymmetryExtension}

In light of these results it is natural to ask if the obstructions discussed above are the only impedance to the existence of a gapped symmetry preserving phase carrying a given anomaly $\omega(A)$ for a discrete global symmetry.  We do not know for certain whether this is so, however it is likely that there are further obstructions.  One possible direction is to use unitarity to constrain the partition function of a TQFT on manifolds like $K3$ which are not null-bordant (i.e.\ they are not the boundary of higher-dimensional manifold).  We comment further on this in section  \ref{posgen} below.

We also note that our results are in agreement with the symmetry extension methods for constructing gapped symmetry preserving phases introduced in \cite{Wang:2017loc} and further studied and applied in \cite{Tachikawa:2017gyf, Wan:2018djl, Wan:2019oyr, Kobayashi:2019lep}.We review these constructions in appendix \ref{apsymext}.  In all examples we have checked where our obstruction applies we find that symmetry preserving gapped states constructed by symmetry extension are indeed impossible. 

\section{Mapping Tori Obstructions }
\label{sec:symobs}

In this section we derive our \MRtext/ regarding discrete global symmetries.  Our principal technical tool is to establish the existence of certain boundary conditions for symmetry defects in unitary TQFTs.  Our line of argument is by contradiction.  We assume that a given anomaly is present in a symmetry preserving TQFT and then derive a contradiction using the boundary conditions.  

Throughout the arguments in this section we assume that the $d$-dimensional TQFT has a unique vacuum on $S^{d-1}.$ This is a technical assumption made for convenience.  In fact, in the statement of our  \MRtext/, multiple vacua on $S^{d-1}$ are allowed as long as the relevant global symmetries are preserved.  The fact that it is sufficient to consider a TQFT with a unique vacuum is because we can project a TQFT onto a given vacuum without changing its anomaly for the symmetries preserved by that state \cite{Sawin:1995rh}.  This projected theory then behaves as if it has a unique vacuum.\footnote{The projection is technically achieved by constructing the idempotent basis of local operators and inserting one of them into every correlator.}

\subsection{Boundary Conditions for Symmetry Operators}
\label{bcsec}

In this section we derive the key result that for $p\leq 1,$ any $p$-form symmetry operators which are not spontaneously broken in a TQFT necessarily admit boundary conditions.  Throughout, we let $g\in G^{(p)}$ be any $p$-form global symmetry of a TQFT, and $U_g$ be the corresponding $(d-p-1)$-dimensional topological operator.  (Below we sometimes say that $U_g$ is \emph{invertible} to emphasize the fact that there is necessarily an inverse operator $U_{g^{-1}}$.)  

Since we are interested in symmetry preserving TQFTs, we would like to understand more explicitly what it means to say that $g$ is spontaneously broken.  In a topological field theory, a symmetry operator  $U_g$ is spontaneously broken if and only if there exists any $p$-dimensional operator $\cO$ that is charged under $U_g$.  Notice that it is crucial for this statement that our theory is topological.  For instance, in a non-topological field theory an ordinary global symmetry is unbroken if and only if the vacuum is uncharged, but excited states may still be acted upon non-trivially.  By contrast, in a TQFT the only states in the spectrum are at zero energy and therefore the existence of any charged states means the symmetry is spontaneously broken.  

In more detail, the charge of a $p$-dimensional operator $\cO$ under a $p$-form global symmetry with symmetry defect $U_{g}$ can be detected by a correlation function where the operators in question are linked (See Figure \ref{fig:p-form_action}.)  For instance, in the case where spacetime is a $d$-dimensional sphere $S^{d}$ we can place the operators on a Hopf link of $S^{p}$ and $S^{d-p-1}$ in $S^d$ defined as follows.  $S^d$ can be constructed by gluing the open manifolds $M_1 = D^{d-p}\times S^{p}$ and $M_2 = S^{d-p-1}\times D^{p+1}$ along the boundaries $S^{d-p-1}\times S^{p}$. The Hopf link $L_p \sqcup L'_{d-p-1}$ is defined by $L_p = \{0\} \times S^p \in M_1 $ and $L'_{d-p-1} = S^{d-p-1}\times \{0\} \in M_2$ embedded into $S^d$ via the gluing $S^d = M_1 \cup_{S^{p}\times S^{d-p-1}} M_2$.

In general, we call a $p$-dimensional operator $\cO_{p}$ nontrivially linked with another operator $\cO'_{d-p-1}$ if the correlation function $\braket{\cO_p(L_p)\cO'_{d-p-1}(L'_{d-p-1})}$, where the operators are on the Hopf link, differs from the similar correlator where $\cO_p$ and $\cO'_{d-p-1}$ are put on a trivial link inside $S^{d}$.  The correlation function $\braket{\cO_p(L_p)\cO'_{d-p-1}(L'_{d-p-1})}$ can also be regarded as an inner product $\braket{\cO_p|\cO'_{d-p-1}}$ where the states $\ket{\cO_p}$ and $\ket{\cO_{d-p-1}'}$ in the Hilbert space on $S^{p}\times S^{d-p-1}$ are defined by the open geometries $M_1$ and $M_2$ with corresponding operator insertions on $L_p\in M_1$ and $L'_{d-p-1}\in M_2$.  We also write these states as 
\begin{equation}
\ket{\cO_k}=Z[D^{d-k}\times S^k_{\cO_k}]~, \hspace{.5in}\ket{\cO_{d-k-1}}=Z[S^{d-k-1}_{\cO_{d-k-1}}\times D^{k+1}]~,
\end{equation}
where $S^p_{\cO_p}$ denotes the sphere wrapped by the operator $\cO_p$, and localized at a point in the transverse direction, and $Z[M]$ denotes the state in the Hilbert space on $\partial M.$ 

With these conventions we can define the linking number of the operators $\cO_{k}$ and $\cO'_{d-k-1}$ by \begin{equation}\label{linkdef}
	\mathrm{link}(\cO_k,\cO'_{d-k-1}) := \frac{\braket{\cO_k|\cO'_{d-k-1}}\braket{1_k|1_{d-k-1}}}{\braket{\cO_k|1_{d-k-1}}\braket{1_k|\cO'_{d-k-1}}}~,
\end{equation}
where $1_k$ denotes the trivial $k$-dimensional operator.  Applied to $O'_{d-p-1}=U_{g},$ a $p$-form symmetry operator, the linking number is the representation of $g$ acting on the linked operator $\cO_{p}$. Thus, a symmetry operator $U_{g}$ of a $p$-form global symmetry is unbroken if and only if, for all operators $\cO_{p}$ of dimension $p$, the associated linking number $\mathrm{link}(\cO_p,U_{g})$ is one.  

Before proceeding further, we need the following lemma and its corollary:
\begin{lem}\label{lem:ZXd}
	Define the double $\Delta X^d$ of an open $d$-dimensional manifold $X^d$ to be $\Delta X^d \equiv \overline{X^d}\cup_{\partial X^d} X^d$ where $\overline{X^d}$ is the orientation reversal of $X^d$. Then, if $X^d$ can be embedded into the sphere $S^d$, the partition function of a unitary (reflection-positive) $d$-dimensional TQFT on $\Delta X^d$ is real and positive:
	\begin{equation}
		Z[\Delta X^d]> 0~.
	\end{equation}
\end{lem}
To establish this, note that the TQFT on the open manifold $X^d$ defines a state $\ket{X^d}$ in the Hilbert space on $ \partial X^d$, and the partition function $Z[\Delta X^d]$ is equal to the norm $\braket{X^d|X^d}$ of the state, which is necessarily real and non-negative. To show that it is non-zero, we observe that since $X^d$ can be embedded into $S^d$ the partition function $Z[S^d]$ on the sphere is an inner product of $\ket{X^d}$ and another state. However, $Z[S^d]$, is also the norm of the vacuum on $S^{d-1}$ (associated to the identity local operator) and is strictly greater than $0$.  Thus we conclude that the state $\ket{X^d}$ is non-zero and hence $Z[\Delta X^d] = \braket{ X^d| X^d}>0$.
Applying this to the case with $X^d = D^{k}\times S^{d-k}$, we get the corollary:
\begin{cor}\label{cor:ZSS}
	The partition function of a unitary (reflection-positive) $d$-dimensional TQFT on $S^k\times S^{d-k}$ is real and positive:
	\begin{equation}
		Z[S^k\times S^{d-k}] > 0~.
	\end{equation}
\end{cor}

We now establish a crucial fact regarding boundary conditions of symmetry operators.  Specifically we establish that for $p\leq 1$, a $p$-form symmetry is preserved in a TQFT, if and only if it is possible to cut the symmetry defect open using a boundary condition.  Notice that one direction of this statement is obvious: if we can cut the symmetry defect open without modifying correlation functions, then it must be the case that all operators are uncharged, since by cutting we can change any link to the trivial link.  Thus, the interesting statement is the converse.

\begin{prop}\label{prop:openhall}
	Let $g\in G^{(p)}$ be an element of the $p$-form symmetry of a TQFT with a unique vacuum on $S^{d-1}$, and suppose that $g$ is not spontaneously broken.  If $g \neq 1\in G^{(p)}$ we further assume that $p\leq 1.$ Then there exists a boundary condition $\phi$ for the corresponding $(d-p-1)$-dimensional symmetry operator $U_{g}$ such that: 
	\begin{equation}
		\braket{U_g(S^{d-p-1}\times \{\pt\})\cdots }_{M^d} = \braket{U_g(D^{d-p-1}_\phi\times \{\pt\})\cdots }_{M^d}~,
		\label{eq:openhall}
	\end{equation}
where $M^d$ is a $d$-dimensional manifold, $D^{d-p-1}_\phi$ is a ball inside $S^{d-p-1}$ with boundary condition $\phi$, and $\cdots$ represents any other insertions of operators that do not intersect with $S^{d-p-1}\setminus D^{d-p-1}_\phi$.
\end{prop}
\begin{remark}
	The boundary condition $\phi$ constructed by proposition \eqref{prop:openhall} need not be fully local, i.e.\ it need not satisfy the (general version of) the Cardy conditions,\footnote{The Cardy conditions for 2d a TQFT are described in \cite{Moore:2006dw}. Boundary conditions for general 3d TQFTs are studied in e.g.\ \cite{Fuchs:2012dt}.} which are not needed for our argument below.  Thus it may be more appropriate to think of $\phi$ as a ``boundary wave function."  We slightly abuse terminology throughout and refer to $\phi$ as a boundary condition with this caveat in mind.
\end{remark}

To prove the proposition, we first observe that since the $p$-form symmetry element $g$ is not spontaneously broken we know that for all $p$-dimensional operators $\cO_{p}$, $\mathrm{link}(\cO_p,U_{g})$ must be 1.\footnote{This does not necessarily mean that $\cO_p$ is trivial. A counter example is a one-form symmetry in the 3+1d $\mathbb{Z}_N\times \mathbb{Z}_N$ gauge theory; see \cite{Hsin:2019fhf}. }  Let us translate this to a condition on states in the Hilbert space on $S^{d-p-1}\times S^{p}$ using the discussion around \eqref{linkdef}.

The key point is that, when $p\le 1$, the Hilbert space on $S^{p}\times S^{d-p-1}$ is spanned by the states of the form $\ket{\cO_{p}}$, where the $\mathcal{O}_{i}'s$ range over all operators in the theory. To see this, note that the $p$-dimensional operators $\cO_p$ are in one-to-one correspondence with the boundary conditions of the $p+1$-dimensional TQFT obtained by compactifying the original TQFT on $S^{d-p-1}$. The correspondence is by taking the radial coordinate of $\cO_p$. Indeed this should be regarded as the definition of extended operators in the TQFT, see, for example, \cite{Kapustin:2010ta}.
When $p=0$, any state of the 1d TQFT defines the boundary condition and in turn a local operator of the ambient $d$-dimensional TQFT.
When $p=1$, we can use the classification theorem\cite{Moore:2006dw} of the boundary conditions of 2d TQFT, which in particular implies that the Hilbert space associated with $S^1$ of a $2d$ (unitary) TQFT is spanned by boundary states of boundary conditions.\footnote{The theorem of \cite{Moore:2006dw} assumes the technical condition that the Frobenius algebra associated with the closed sector of the given $2d$ TQFT is semi-simple. Under the assumption of unitarity, this condition is automatically satisfied \cite{Durhuus:1993cq,Sawin:1995rh}.}

Given that $\ket{\cO_p}$ spans the Hilbert space, the fact that $U_{g}$ always produces trivial linking number means that as a state:
\begin{equation}\label{ugis}
\ket{U_{g}}=\ket{1_{d-p-1}}~.
\end{equation}
In particular, we note that the state $|U_{g} \rangle$ cannot vanish since $\braket{1_p|1_{d-p-1}} = Z[S^d] >0$  by reflection-positivity and the fact that the vacuum on $S^{d-1}$ is a nonzero state (See also Lemma \ref{lem:ZXd}.).   This means that the correlator $\braket{U_g(S^{d-p-1}\times \{\pt\})}_{M^d}=\braket{U_{g}|1_{d-p-1}}$ is equal to the norm that is strictly positive from Corollary \ref{cor:ZSS}:
\begin{equation}\label{eq:Ugpos}
	\braket{U_g(S^{d-p-1}\times \{\pt\})}_{S^{d-p-1}\times S^{p+1}}=\braket{U_{g}|1_{d-p-1}}=\braket{1_{d-p-1}|1_{d-p-1}} = Z[S^{d-p-1}\times S^{p+1}]>0~.
\end{equation}

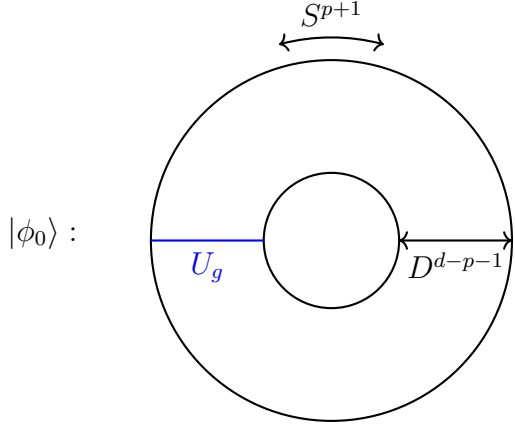
\begin{figure}[t]
	\centering
	$\ket{\phi_0}:\quad\quad$
	\begin{tikzpicture}[scale=.6,thick,baseline = 0]
		\draw circle(4);
		\draw circle(1.5);
		\draw[blue] (-4,0)--node[midway,anchor=north] {$U_g$}(-1.5,0);
		\draw[<->] (4,0)-- node[midway,anchor=north] {$D^{d-p-1}$}(1.5,0);
		\draw[<->] (75:4.5) arc(75:105:4.5) node[midway,anchor=south] {$S^{p+1}$};
	\end{tikzpicture}
	\caption{The geometry defining the state $\ket{\phi_0}$ in the Hilbert space $\cH$ on $S^{p+1}\times S^{d-p-2}_{U_g}$. The picture is on the special case of $p=0$ and $d=2$, or can be thought as depicting a 2-dimensional slice of a more general case.}
	\label{fig:phi0}
\end{figure}

The above result can also be understood as the existence of non-trivial states in a symmetry twisted Hilbert space.
Indeed, examine the Hilbert space $\cH$ on $S^{d-p-2}_{U_g}\times S^{p+1}$ where the symmetry defect $U_{g}$ is inserted along $S^{d-p-2}\times \{\pt\}$ and extends along time.
Such states can be viewed as wave functions defined by an open manifold $\ket{\phi_0} := Z[D^{d-p-1}_{U_{g}}\times S^{p+1}]$ where the defect wraps the disk $D^{d-p-1}$. (See Figure \ref{fig:phi0}.)
Using the result \eqref{eq:Ugpos} we can also see that this Hilbert space is non-empty.  We simply take the positive correlator $\braket{U_g(S^{d-p-1}\times \{\pt\})}_{M^d}$, and cut it along an equator in $S^{d-p-1}$.
In this way, we can interpret this correlator as the norm of a non-zero state $\ket{\phi_0}:= Z[D^{d-p-1}_{U_g}\times S^{p+1}]$.

\begin{figure}[t]
	\centering
	$\ket{0}:\quad\quad$
	\begin{tikzpicture}[scale=.6,thick,baseline = 0]
		\draw circle(3);
		\draw[blue] (-3,0)--node[midway,anchor=north] {$U_g$}(3,0);
		\node[anchor=south,blue] at (0,0) {$D^{d-p-1}$};
		\node at (0,3.8) {$D^d$};
	\end{tikzpicture}
	\caption{The geometry defining the state $\ket{0}$ in the Hilbert space $\widetilde{\cH}$ on $S^{d-1}$ with $U_g$ inserted along $S^{d-p-2}$.}
	\label{fig:ket0}
\end{figure}
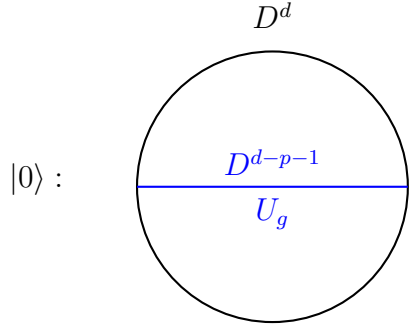
These remarks are now sufficient to establish the existence of a boundary condition for the symmetry defect satisfying the property \eqref{eq:openhall}. 
Let $\ket{0}= Z[D^{d},U_g(\Sigma)]$, where $\Sigma$ is a $D^{d-p-1}$ embedded in $D^{d}$ so that the boundary of the former is embedded in the boundary of the latter. (See Figure \ref{fig:ket0}.)
This state $\ket{0}$ is in the Hilbert space $\widetilde{\cH}$ on $S^{d-1}$ with $U_g$ insertion along $\partial \Sigma \cong S^{d-p-2}$ embedded in $S^{d-1}$. Because the $U_g$ insertion is null-homologus, the Hilbert space is isomorphic to the Hilbert space on $S^{d-1}$ without insertion, which is one-dimensional associated to the unique unit local operator: $\widetilde{\cH}\cong \mathbb{C}$.

Next, we construct a linear map $f: \cH \to \widetilde{\cH}$, by studying the TQFT on a manifold with two boundary components: one an $S^{d-1}$ (related to $\widetilde{\cH}$), and the other an $S^{p+1}\times S^{d-p-2}$ (related to $\cH$).  The appropriate $d$-manifold, $M$, for the construction is a disk $D^{d}$ minus a $D^{p+2}\times S^{d-p-2}$, where the space $D^{p+2}\times S^{d-p-2}$ is the tubular neighborhood of a sphere $S^{d-p-2}$ in the interior of the disk.  We also define $\widetilde{\Sigma}\in M$ as $\widetilde{\Sigma}\cong [0,1]\times S^{d-p-2}$, one of whose boundary components lies in the $S^{d-1}$ at the edge of the disk, and the other lies in the boundary of the tubular neighborhood $D^{p+2}\times S^{d-p-2}$. (See Figure \ref{fig:mapf}.)  The TQFT on $M$ with $U_g$ inserted along $\widetilde{\Sigma}$ defines the desired map $f: \cH \to \widetilde{\cH}$. 

To find a boundary condition with the property \eqref{eq:openhall}, it is now sufficient to find a state $\ket{\phi} \in \cH$ satisfying 
\begin{equation}
	f\ket{\phi} = \ket{0}~.
	\label{eq:ketphiprop}
\end{equation}
A pictorial proof of \eqref{eq:openhall} given a state $\ket{\phi}$ with property \eqref{eq:ketphiprop} is given in Figure~\ref{fig:openhall}.
\begin{figure}[t]
	\centering
	$f:\quad\quad$
	\begin{tikzpicture}[scale=.6,thick,baseline = 0]
		\draw circle(4);
		\draw (1.5,0) circle(1);
		\draw (-1.5,0) circle(1);
		\node at (0,4.8) {$D^d$};
		\node at (0,1.8) {$S^{p+1}\times S^{d-p-2}$};
		\draw[blue] (-4,0) -- node[midway,anchor = north] {$U_g$} (-2.5,0);
		\draw[blue] (4,0) -- (2.5,0);
	\end{tikzpicture}
	\caption{The geometry defining the linear map $f$ from $\cH$ to $\widetilde{\cH}$.}
	\label{fig:mapf}
\end{figure}
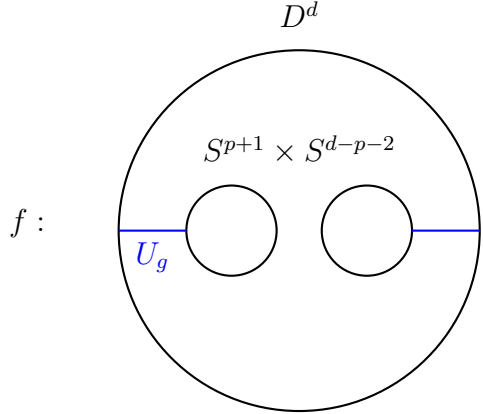
\begin{figure}
	\centering
	\subcaptionbox{}[.2\linewidth]{
		\begin{tikzpicture}[scale = .6,thick,baseline = 0]
			\draw[blue] (-2.5,0) -- (2.5,0);
			\node[blue,anchor=south] at (2,0) {$U_g$};
			\fill[fill=none] circle(1.5);
		\end{tikzpicture}
	}
	\subcaptionbox{}[.2\linewidth]{
		\begin{tikzpicture}[scale = .6,thick,baseline = 0]
			\draw[blue] (-2.5,0) -- (2.5,0);
			\draw[fill=white] circle(1.5);
			\node at (0,2.1) {$\ket{0}$};
			\node[blue,anchor=south] at (2,0) {$U_g$};
		\end{tikzpicture}
	}
	\subcaptionbox{}[.2\linewidth]{
		\begin{tikzpicture}[scale = .6,thick,baseline = 0]
			\draw[blue] (-2.5,0) -- (2.5,0);
			\draw[fill=white] circle(1.5);
			\draw (-.75,0) circle(.4);
			\draw (.75,0) circle(.4);
			\node at (0,2.1) {$f\ket{\phi} = \ket{0}$};
			\node at (0,.9) {$\ket{\phi}$};
			\draw[blue] (-1.5,0) -- (-1.15,0);
			\draw[blue] (1.5,0) -- (1.15,0);
			\node[blue,anchor=south] at (2,0) {$U_g$};
		\end{tikzpicture}
	}
	\subcaptionbox{}[.2\linewidth]{
		\begin{tikzpicture}[scale = .6,thick,baseline = 0]
			\draw[blue] (-2.5,0) -- (2.5,0);
			\fill[fill=white] circle(1.5);
			\draw (-.75,0) circle(.4);
			\draw (.75,0) circle(.4);
			\node at (0,.9) {$\ket{\phi}$};
			\draw[blue] (-1.5,0) -- (-1.15,0);
			\draw[blue] (1.5,0) -- (1.15,0);
			\node[blue,anchor=south] at (2,0) {$U_g$};
		\end{tikzpicture}
	}
	\caption{Opening a hole on the symmetry operator $U_g$ using the state $\ket{\phi}$ satisfying \eqref{eq:ketphiprop}. (a): a local patch around a point on the operator $U_g$. (b): The radial time evolution around the selected point on $U_g$ defines the state $\ket{0}$. (c): Given a state $\ket{\phi} \in \cH$ with the property \eqref{eq:ketphiprop}, we can fill the sphere $S^{d-1}$ on which $\ket{0}$ is defined with the geometry associated with the map $f$. (d): The procedure effectively opens up a hole on the defect $U_g$. }
	\label{fig:openhall}
\end{figure}
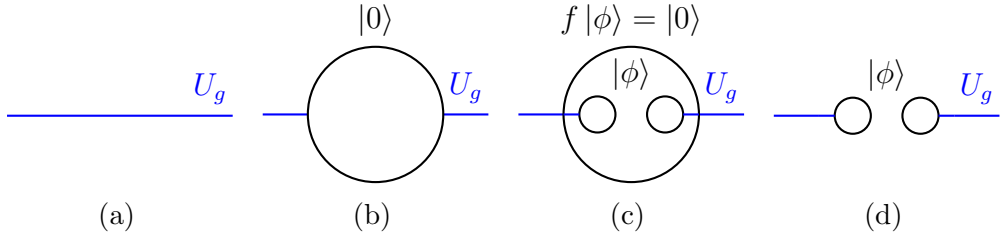
\begin{figure}[t]
	\centering
	$f\!\ket{\phi_0}:\quad\quad$
	\begin{tikzpicture}[scale=.5,thick,baseline = 0]
		\draw circle(4);
		\draw (-2,.5) ..controls (0,1.5) .. (2,.5);
		\draw (-1.5,0) ..controls (0,.7) .. (1.5,0);
		\draw[blue] (-4,0) --  (-2.8,0) .. controls (-1.7,0) and (-.7,.9) .. (0,.9) .. controls (.7,.9) and (1.7,0) .. (2.8,0) -- (4,0);
		\node[anchor = north,blue] at (-3,0) {$U_g$};
	\end{tikzpicture}
	\caption{The geometry corresponding to the state $f\ket{\phi_0}$, construced by gluing the geometry depicted in Figure~\ref{fig:phi0} with the geometry in Figure~\ref{fig:mapf}. The handle $D^{d-p-1}\times S^{p+1}$ is attached to the disk $D^d$ and the operator $U_g$ goes through the handle.}
	\label{fig:fphi0}
\end{figure}
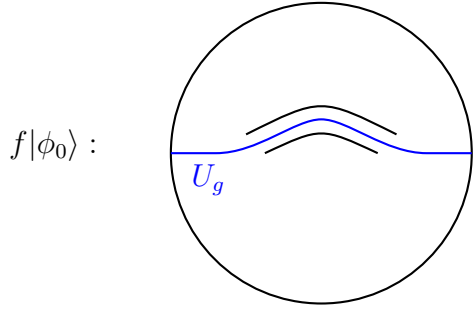
As $\widetilde{\cH}\cong \mathbb{C}$, we can define such a state by
\begin{equation}
	\ket{\phi} = \frac{\braket{0|0}}{\braket{0|f|\phi_0}}\ket{\phi_0}~,
\end{equation}
if we can show that $\braket{0|f|\phi_0}\neq 0$.
The state $f\!\ket{\phi_0}$ corresponds to the geometry depicted in Figure~\ref{fig:fphi0}, which is the disk with the handle $D^{d-p-1}\times S^{p+1}$ attached.
Then, the geometry of $\braket{0|f|\phi_0}$ is the handle attached sphere, which is diffeomorphic to $S^{d-p-1}\times S^{p+1}$.\footnote{This kind of operation on the geometry is called surgery. A recent study on surgery in TQFT, see, e.g.\ \cite{Wang:2019diz}.}
The $U_g$ insertion can also be tracked and it wraps $S^{d-p-1}$ in $S^{d-p-1}\times S^{p+1}$.
Hence, we have $\braket{0|f|\phi_0} = \braket{U_g(S^{d-p-1}\times \{\pt\})}_{S^{d-p-1}\times S^{p+1}}$,
and \eqref{eq:Ugpos} ensures that 
$\braket{0|f|\phi_0}\neq 0$.

\subsection{Mapping Torus Obstructions}\label{sec:MappingTorus}

We now prepare to establish \MRtext/.  Since this statement involves mapping torus manifolds let us first explain why such manifolds arise naturally in the study of anomalies.  

Consider a $d$-dimensional spacetime manifold $M$, and let $N$ be the $(d+1)$-manifold $M\times I$ where $I$ is an interval.  The boundary of $N$ has two components, each identified with $M$.  Let $A_{i}$ for $i=1,2$ be background fields on the two copies of $M$, and let $A$ be an extension of $A_{i}$ to $N$.  Then following the discussion in section \ref{sec:'thooftrev} the partition function:
\begin{equation}
Z[A_{1}]\exp\left(2\pi i \int_N \omega(A)\right)Z^{*}[A_{2}]
\end{equation}
is gauge invariant with respect to any gauge parameter defined on $N$.  

In particular, we can apply this observation to the situation where $A_{2}$ differs from $A_{1}$ merely by a gauge transformation  on $M$.  Applying such a gauge transformation and taking $A$ to be constant along $N$, (so that $\omega$ does not contribute) we learn:
\begin{equation}\label{inttrans}
	Z[A_1]\exp\left(2\pi i \int_N \omega(A)\right)Z[A_1^\lambda]^{*} = Z[A_1]Z[A_1]^*~.
\end{equation}
In particular since $\omega$ is real we see $|Z[A^\lambda] |=|Z[A]|$.  Thus \eqref{inttrans} implies the anomalous transformation law:
\begin{equation}
Z[A_1]\exp\left(2\pi i \int_N \omega(A)\right)=Z[A_1^{\lambda}]~.
\end{equation}
Because the background $A$ on $N$ interpolates between configurations on $M$ that differ only by a gauge transformation, the phase $\exp\left(2\pi i \int_N \omega(A)\right)$ can be regarded as the evaluation of $\omega$ on the closed manifold obtained by gluing the two ends of $N$ (perhaps using a diffeomorphism of $M$), which is the mapping torus, and with the background given by gluing $A$ with the gauge transformation $\lambda$.

We now establish our \MRtext/. We assume the setup and notation of the original statement.  Using the observations above, a discrete anomaly that is activated on a mapping torus may be equivalently reinterpreted in terms of a simple anomalous transformation law of the $d$-dimensional partition function on  $S^{p_{1}+1}\times S^{p_{2}+1},$ with $p_{1}+p_{2}=d-2$.  Specifically:
\begin{eqnarray}
		Z[N^d,(A_1,A_2,0,0,0)] &=& \exp\left(2\pi i\int_{M^{d+1}}\omega(A)\right)Z[f(N^d),(f(A_1),f(A_2),\mathrm{d}\lambda,\mathrm{d}\Lambda_1,\mathrm{d}\Lambda_2)] \nonumber \\
		 &=& \exp\left(2\pi i\int_{M^{d+1}}\omega(A)\right)Z[N^d,(A_1,A_2,\mathrm{d}\lambda,\mathrm{d}\Lambda_1,\mathrm{d}\Lambda_2)]~, \label{eq:Zanomalous}
\end{eqnarray}
where $N^d= S^{p_1+1}\times S^{p_2+1}$, $Z[N^d,A=(A_1,A_2,B_0,B_1,B_2)]$ denotes the partition function on $N^d$ with the background $A$. The backgrounds $A_1$, $A_2$ are defined in our \MRtext/ and wrap the spheres $S^{p_{i}+1}$.   Meanwhile, the 
background  fields $B_{i}$ are trivial in $d$ dimensions, but define non-trivial classes on the $(d+1)$-dimensional mapping torus.  Specifically, they are the backgrounds that interpolate between zero and $d\Lambda$, where $(\lambda,\Lambda_1,\Lambda_2)$ are the gauge transformation parameters for the symmetries $G^{(0)},G^{(p_1+1)}, G^{(p_2+1)}.$   To reproduce the setup of \MRtext/, we take these parameters to be constant $\lambda = g_0[\pt]$, $\Lambda_1=g_3[S^{p_1+1}]$, and $\Lambda_2 = g_4[S^{p_2+1}]$.

Now we simply observe that $\mathrm{d}\Lambda = 0$, and therefore  the transformation law \eqref{eq:Zanomalous} implies:
\begin{equation}\label{orstate}
\exp\left(2\pi i \int_{M^{d+1}}\omega(A)\right)=1~, \hspace{.25in}\mathrm{or}\hspace{.25in}Z[N^d,(A_1,A_2)]=0~.
\end{equation}
Therefore, to prove our \MRtext/, it suffices to show the following proposition (recall from the preamble to section \ref{sec:symobs} that there is no loss in generality in assuming a unique vacuum on $S^{d-1}$):
\begin{prop}\label{prop:ZSS}
	A unitary TQFT with a unique vacuum preserving the $G^{(p_{1})}\times G^{(p_{2})}$ symmetry satisfies
	\begin{equation}
		Z[N^d,(A_1,A_2)]\neq 0,
	\end{equation}
	where $N^d = S^{p_1+1}\times S^{p_2+1}$, $A_1 = g_1 [S^{p_1+1}]$, $A_2=g_2[S^{p_2+1}]$ with $g_1\in G^{(p_1)}$, $g_2\in G^{(p_2)}$.
\end{prop}

We can establish proposition $\ref{prop:ZSS}$ as follows. (Below we assume $p_2> 0$. The case $p_2=0$ case can be proven by just ignoring the associated operator in the following argument.)  The partition function $Z[N^d,(A_1,A_2)]$ is equivalent to the correlation function
	\begin{equation}
		Z[N^d,(A_1,A_2)] = \langle U_{g_1}(S^{p_1+1}\times\{\pt_2\}) U_{g_2}(\{\pt_1\} \times S^{p_2+1}) \rangle_{N^d}~,
		\label{eq:ZNd}
	\end{equation}
	where $\pt_1\in S^{p_1+1}$ and $\pt_2\in S^{p_2+1}$ are points,  and $U_{g_1},U_{g_2}$ are the symmetry operators corresponding to $g_1\in G^{(p_1)}$ and $g_2\in G^{(p_2)}$.  
	
By assumption, the symmetries $G^{(p_1)}$ and $G^{(p_2)}$ are preserved and therefore, we may use proposition \ref{prop:openhall} on each such operator to cut them open using an appropriate boundary condition.  	This in turn means that 
\begin{equation}
Z[N^d,(A_1,A_2)] =	\langle U_{g_1}(U_1\times\{\pt_2\})U_{g_2}(\{\pt_1\}\times  U_2)\rangle_{N^d}~,
		\label{eq:disks}
\end{equation}
where $U_i$ is a small ball around $\pt_i\in S^{p_i+1}$.   

The  symmetry defects are now concentrated in an arbitrarily small region $U_1\times U_2$ near their intersection point in $N^{d}$.  We can therefore surround them will a small ball $U$, and replace the symmetry defects by a state in the Hilbert space on $\partial U\cong S^{d-1}$.  Since such states are equivalent to local operators, we are effective replacing the effect of the symmetry defects by the insertion of a local operator $X$ at $(\pt_1,\pt_2)$.\footnote{Since $X$ arises from the intersection of symmetry defects, its precise definition depends on the choice of local counterterms.  Such choices can modify the phase of $X$ but not its absolute value.}

Now we use the assumption that the TQFT has a unique vacuum.  This means that the operator $X$ is proportional to the unique local operator in the theory, namely the identity $\mathds{1}$ as $X = c_X\mathds{1}$ with a number $c_X$. Let $\tilde{X}$ be the similar operator to $X$ obtained by replacing $U_{g_2}$ by $U_{g_2^{-1}}$ in the definition of $X$. Then, we can prove that $X\tilde{X}\neq 0$, in particular $c_X\neq 0$ by carefully colliding the two local operators as illustrated in Figure~\ref{fig:XXtil}. 
\tikzset{->-/.style={decoration={
  markings,
  mark=at position #1 with {\arrow{>}}},postaction={decorate}}}
\tikzset{-<-/.style={decoration={
  markings,
  mark=at position #1 with {\arrow{<}}},postaction={decorate}}}
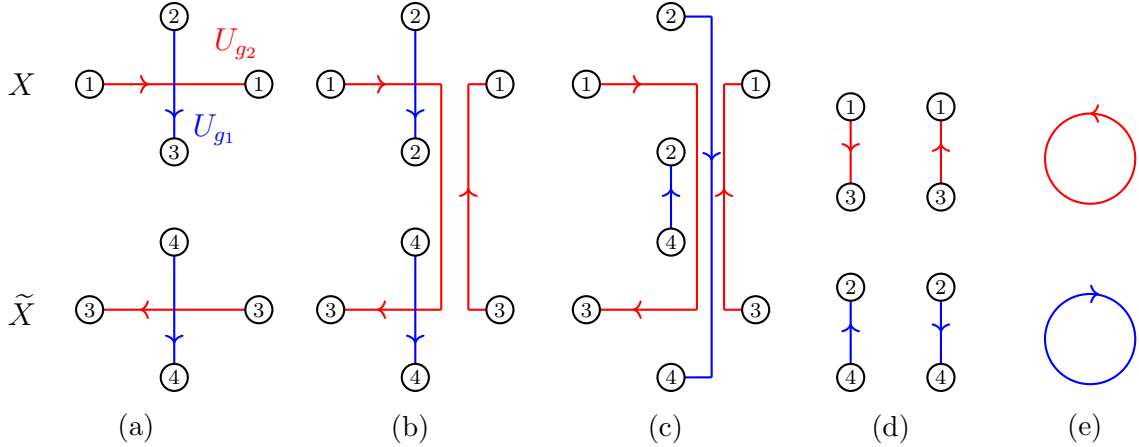
\begin{figure}[t]
	\centering
	\subcaptionbox{}[.23\linewidth]{
		\begin{tikzpicture}[scale = .6,thick,baseline = 0]
			\draw[->-=.35,red] (0,0) coordinate (a1) --++(3.75,0) coordinate (a3);
			\draw[fill=white] (a1) circle(.3);
			\node at (a1) {\scriptsize 1};
			\draw[fill=white] (a3) circle(.3);
			\node at (a3) {\scriptsize 1};
			\draw[->-=.75,blue] ($(a1)!.5!(a3)$) ++ (0,1.5) coordinate(b1) --++(0,-3) coordinate(b3);
			\draw[fill=white] (b1) circle(.3);
			\node at (b1) {\scriptsize 2};
			\draw[fill=white] (b3) circle(.3);
			\node at (b3) {\scriptsize 3};
			\node[red] at ($(a3)+(-.5,.9)$) {$U_{g_2}$};
			\node[blue] at ($(b3)+(.9,.5)$) {$U_{g_1}$};
			\draw[-<-=.35,red] (0,-5) coordinate (c1) --++(3.75,0) coordinate (c3);
			\draw[fill=white] (c1) circle(.3);
			\node at (c1) {\scriptsize 3};
			\draw[fill=white] (c3) circle(.3);
			\node at (c3) {\scriptsize 3};
			\draw[->-=.75,blue] ($(c1)!.5!(c3)$) ++ (0,1.5) coordinate(d1) --++(0,-3) coordinate(d3);
			\draw[fill=white] (d1) circle(.3);
			\node at (d1) {\scriptsize 4};
			\draw[fill=white] (d3) circle(.3);
			\node at (d3) {\scriptsize 4};
			\node at ($(a1)+(-1.5,0)$) {$X$};
			\node at ($(c1)+(-1.5,0)$) {$\widetilde{X}$};
		\end{tikzpicture}
	}
	\subcaptionbox{}[.2\linewidth]{
		\begin{tikzpicture}[scale = .6,thick,baseline = 0]
			\draw[->-=.5,red] (0,0) coordinate (a1) --++(2.45,0) coordinate (a4);
			\draw[red] (a4) ++ (.6,0) coordinate(a5) --++(.7,0) coordinate(a3);
			\draw[fill=white] (a1) circle(.3);
			\node at (a1) {\scriptsize 1};
			\draw[fill=white] (a3) circle(.3);
			\node at (a3) {\scriptsize 1};
			\draw[->-=.75,blue] ($(a1)!.5!(a3)$) ++ (0,1.5) coordinate(b1) --++(0,-3) coordinate(b3);
			\draw[fill=white] (b1) circle(.3);
			\node at (b1) {\scriptsize 2};
			\draw[fill=white] (b3) circle(.3);
			\node at (b3) {\scriptsize 2};
			\draw[-<-=.5,red] (0,-5) coordinate (c1) --++(2.45,0) coordinate (c4);
			\draw[red] (c4) ++ (.6,0) coordinate(c5) --++(.7,0) coordinate(c3);
			\draw[red] (a4) -- (c4);
			\draw[red,-<-=.5] (a5) -- (c5);
			\draw[fill=white] (c1) circle(.3);
			\node at (c1) {\scriptsize 3};
			\draw[fill=white] (c3) circle(.3);
			\node at (c3) {\scriptsize 3};
			\draw[->-=.75,blue] ($(c1)!.5!(c3)$) ++ (0,1.5) coordinate(d1) --++(0,-3) coordinate(d3);
			\draw[fill=white] (d1) circle(.3);
			\node at (d1) {\scriptsize 4};
			\draw[fill=white] (d3) circle(.3);
			\node at (d3) {\scriptsize 4};
		\end{tikzpicture}
	}
	\subcaptionbox{}[.2\linewidth]{
		\begin{tikzpicture}[scale = .6,thick,baseline = 0]
			\draw[->-=.5,red] (0,0) coordinate (a1) --++(2.45,0) coordinate (a4);
			\draw[red] (a4) ++ (.6,0) coordinate(a5) --++(.7,0) coordinate(a3);
			\draw[fill=white] (a1) circle(.3);
			\node at (a1) {\scriptsize 1};
			\draw[fill=white] (a3) circle(.3);
			\node at (a3) {\scriptsize 1};
			\draw[blue] ($(a1)!.5!(a3)$) ++ (0,1.5) coordinate(b1) ++(0,-3) coordinate(b3);
			\draw[-<-=.5,red] (0,-5) coordinate (c1) --++(2.45,0) coordinate (c4);
			\draw[red] (c4) ++ (.6,0) coordinate(c5) --++(.7,0) coordinate(c3);
			\draw[red] (a4) -- (c4);
			\draw[red,-<-=.5] (a5) -- (c5);
			\draw[fill=white] (c1) circle(.3);
			\node at (c1) {\scriptsize 3};
			\draw[fill=white] (c3) circle(.3);
			\node at (c3) {\scriptsize 3};
			\draw[blue] ($(c1)!.5!(c3)$) ++ (0,1.5) coordinate(d1) ++(0,-3) coordinate(d3);
			\draw[-<-=.5,blue] (b3) -- (d1);
			\draw[blue] (b1) -- ++(.9,0) coordinate(e1);
			\draw[blue,->-=.4] (e1) -- ++ (0,-8) coordinate(e2);
			\draw[blue] (e2) -- (d3);
			\draw[fill=white] (d1) circle(.3);
			\node at (d1) {\scriptsize 4};
			\draw[fill=white] (d3) circle(.3);
			\node at (d3) {\scriptsize 4};
			\draw[fill=white] (b1) circle(.3);
			\node at (b1) {\scriptsize 2};
			\draw[fill=white] (b3) circle(.3);
			\node at (b3) {\scriptsize 2};
		\end{tikzpicture}
	}
	\subcaptionbox{}[.15\linewidth]{
		\begin{tikzpicture}[scale = .6,thick,baseline = 0]
			\draw[red,->-=.5] (0,0) coordinate(1-l) --++(0,-2) coordinate(3-l);
			\draw[red,-<-=.5] (2,0) coordinate(1-r) --++(0,-2) coordinate(3-r);
			\draw[blue,-<-=.5] (0,-4) coordinate(2-l) --++(0,-2) coordinate(4-l);
			\draw[blue,->-=.5] (2,-4) coordinate(2-r) --++(0,-2) coordinate(4-r);
			\foreach \x in {1,2,3,4}{
			\draw[fill=white] (\x-l) circle(.3);
			\node at (\x-l) {\scriptsize \x};
			\draw[fill=white] (\x-r) circle(.3);
			\node at (\x-r) {\scriptsize \x};
		}
		\end{tikzpicture}
	}
	\subcaptionbox{}[.15\linewidth]{
		\begin{tikzpicture}[scale = .6,thick,baseline = 0]
			\draw[red,->-=.25] (0,0) circle(1);
			\draw[blue,-<-=.25] (0,-4) circle(1);
			\node at (0,-5) {};
		\end{tikzpicture}
	}
	\caption{The operation showing $X \widetilde{X}\neq0$ and hence $X\neq0$. (a): The definition of $X$ and $\widetilde{X}$ is depicted. $X$ is defined by the intersection of the operator $U_{g_1}$ on $D^{d-p_1-1}$ and the operator $U_{g_2}$ on $D^{d-p_2-1}$. The boundary condition of these operators are the ones constructed in proposition \ref{prop:openhall}. $\widetilde{X}$ involves $U_{g_2^{-1}}$ while $X$ involves $U_{g_2}$ (which is pictorially indicated with the orientation of arrows above). In the picture is a 2-dimensional slice of the two operators. Small circles with the same number in the picture belong to the boundary of the same disk.
		(b): Place $X$ and $\widetilde{X}$ close to each other, and deform the $U_{g_2}$ portion of $X$ and $\tilde{X}$ so that the to operators are connected. Now $U_{g_2}$ is on an annulus $[0,1]\times S^{d-p_2-2}$.
	(c): Slide the two intersection points between $U_{g_1}$ and $U_{g_2}$ so that the pair annihilates. Now $U_{g_1}$ is also on an annulus $[0,1]\times S^{d-p_1-2}$, and $U_{g_1}$ and $U_{g_2}$ do not intersect.
(d): The two annuli $[0,1]\times S^{d-p_2-2}$ and $[0,1]\times S^{d-p_1-2}$ do not link with each other for the dimensional reasons, and thus the two can be untangled in the $d$-dimensional spacetime.
(e): By using proposition \ref{prop:openhall} in reverse, we can add caps to these annuli and we end up with the operators $U_{g_1}$ and $U_{g_2}$ on small spheres, which in turn can be collapsed to the unit operator.}
\label{fig:XXtil}
\end{figure}

Finally, using Corollary \ref{cor:ZSS} with these observations, we conclude
	\begin{equation}
		Z[N^d,(A_1,A_2)] = \langle X \rangle_{N^d} =c_X Z[N^d]\neq 0.
		\label{eq:ZNd}
	\end{equation}
Comparing with \eqref{orstate} proves our \MRtext/.

\subsection{A Possible Generalization Involving $K3$ Manifolds}\label{posgen}

Our result is a necessary condition for the existence of a symmetry preserving gapped state.  However we have no reason to believe that it is sufficient.  Here we describe a possible avenue for generalization.\footnote{After the submission of the preprint version of this paper to arXiv, the authors succeeded in fulfilling the generalization in \cite{Cordova:2019jqi}.}

Let us consider four-dimensional fermionic theories with $\mathbb{Z}_{N}$ global symmetry.  These theories have a possible anomaly $\omega,$ which satisfies 
\begin{equation}\label{omk3}
	e^{2\pi\mathrm{i} \omega(S^1_h\times K3)} \neq 1~,
\end{equation}
where $S^1_h$ denotes the circle with the minimal $\mathbb{Z}_N$ holonomy, the additional factor is the four-manifold $K3$ with no additional background fields.  Moreover, unlike the bosonic anomalies for a finite groups, one cannot construct a symmetry preserving gapped boundary condition for $\omega$ using the the symmetry extension method of \cite{Wang:2017loc}.\footnote{ Indeed, the only necessary background to activate $\omega$ in \eqref{omk3} is the holonomy around $S^1$.  This can always be uplifted into an extended group, and therefore the anomaly $\omega$ cannot be trivialized by any such extension.}  Therefore, it is tempting to expect that no symmetry-preserving TQFT exists.

Applying the logic in the previous subsection, for any QFT with the anomaly $\omega$ we deduce that the partition function on $K3$ vanishes
\begin{equation}
	Z[K3] = 0~.
\end{equation}
However, it is not obvious whether there exists a unitary TQFT with vanishing $Z[K3]$ because Lemma \ref{lem:ZXd} is not applicable to this case, and it would be interesting to try to find an example or to exclude such a possibility.  Note also that the absence of symmetry extensions trivializing the anomaly also implies any $4d$ TQFT carrying the anomaly $\omega$ is necessarily not  a finite (0-form) group gauge theory.

\section{Examples with the Obstruction}\label{disexsec}

In this section we describe examples of theories and anomalies that do not admit symmetry preserving gapped phases due to our \MRtext/.
\subsection{$2d$ Lieb-Schultz-Mattis Theorem}\label{sec:LMS}

As a first example, let us revisit the $2d$ LSM theorem \cite{Lieb:1961fr}.  The data of a non-trivial $2d$ TQFT is a list of topological local operators and their fusion algebra \cite{DijkgraafThesis, Moore:2006dw}. Thus, for instance if we insist on studying $2d$ TQFTs with a unique vacuum it is immediate that all zero-form symmetries are not spontaneously broken and furthermore do not act on the theory.  In particular this means that such symmetries cannot have 't Hooft anomalies.  

To see how our \MRtext/ formally reproduces this claim, consider for simplicity a bosonic theory with zero-form symmetry $\mathbb{Z}_{N}$.  The general anomaly for such a theory takes the form
\begin{equation}\label{lsm}
2\pi i\omega(A)=\frac{2\pi i k}{N} A \cup \beta(A)~.
\end{equation}
Here, $A$ is the $\mathbb{Z}_{N}$ gauge field, and $\beta(A)\in H^{2}(M, \mathbb{Z})$ is the Bockstein operation.  The anomaly is characterized by $k$ which is an integer modulo $N$.

To evaluate our obstruction, we consider a three-manifold defined as a mapping torus of $T^2$ where the twist $f$ is by the $SL(2,\mathbb{Z})$ element (large diffeomorphism)
\begin{equation}
T^N = \begin{pmatrix}1 & N\\ 0 & 1 \end{pmatrix}~.
\end{equation}
On this three-manifold, we take the background $A$ to be the mod $N$ reduction of the $\Z$-valued class that is not invariant under $T^N$ but invariant mod $N$.  Such a class by definition has $\beta(A)$ non-zero and the anomaly \eqref{lsm} evaluates non-trivially.  Therefore we conclude that no symmetry preserving gapped phase exists.

\subsection{Factorized Anomalies of Discrete Gauge Theories}\label{secdiscex}

Consider a topological $\mathbb{Z}_{N}$ gauge theory in $d$ dimensions based on a dynamical $p$-form gauge field $a_{p}$ and its magnetic dual $a_{d-p-1}$.  This theory has two global symmetries, a $(d-p-1)$-form symmetry $\mathbb{Z}_{N}^{(d-p-1)}$ with background field $A_{d-p}$ and a $p$-form symmetry $\mathbb{Z}_{N}^{(p)}$ with background field $A_{p+1}$.  The action coupled to background fields is:
\begin{equation}
S=\frac{2\pi}{N} \int a_{p} \cup da_{d-p-1}+a_{p} \cup A_{d-p}+a_{d-p-1}\cup A_{p+1}~.
\end{equation}  
In this discrete gauge theory, the realization of these symmetries is that they are both spontaneously broken.  Below we will see that in some cases this is due to the anomaly.

These two higher-form symmetries participate in a mixed anomaly characterized by
\begin{equation}\label{factorizedw}
2\pi i\omega=\frac{2\pi i}{N}\int A_{d-p}\cup A_{p+1}~.
\end{equation}
Thus, we would like to understand whether this anomaly alone implies that these symmetries are spontaneously broken.  To answer this we apply our \MRtext/.

The anomaly \eqref{factorizedw} evaluates non-trivially on the $d+1$-manifold $S^{1}\times S^{p+1}\times S^{d-p-1}$ with $A_{p}$ wrapping the cycle $S^{p+1}$ and $A_{d-p}$ wrapping $S^{1}\times S^{d-p-1}$.  Therefore we learn that if $p=0,1$, and theory with anomaly \eqref{factorizedw} necessarily breaks the $p$-form symmetry spontaneously.  Similarly considering the manifold $S^{1}\times S^{p}\times S^{d-p}$ we learn that the $(d-p-1)$-form symmetry must also be broken provided that $d-p-1\leq 1$.  

\subsection{$3d$ Theories with Anomalous One-Form Symmetry}

Next, let us consider $3d$ QFTs with 1-form $\mathbb{Z}_N^{(1)}$ symmetry and anomaly
\begin{equation}
	\tpi\omega = \frac{2\pi\mathrm{i}k}{\gcd(N,2)N}\int \mathfrak{P}(B)~,
\end{equation}
with some coefficient $k$, where $B$ is the $\Z_N^{(1)}$ background and $\mathfrak{P}$ denotes the Pontryagin square. (When $N$ is odd we simply mean the cup-square by the Pontryagin square.)  This anomaly is realized, for instance by $U(1)_{N}$ Chern-Simons theory, where the one-form symmetry is spontaneously broken.  

In \cite{Hsin:2018vcg} it was shown that in a $3d$ TQFT this anomaly is completely captured by the data of the braiding of abelian anyons, and therefore for any $k$ the one-form symmetry is spontaneously broken in any gapped theory. Our \MRtext/ reproduces the same conclusion when $\frac{k}{\gcd(N,2)N}\neq \frac12$, because this anomaly is nontrivial on $S^1 \times S^1 \times S^2$ with background $B=[S^1\times S^1]+[S^2]$.

\subsection{$4d$ Yang-Mills with $\theta = \pi$}

In \cite{Gaiotto:2017yup} it was found that $4d$ pure Yang-Mills theory with gauge group $G$ and $\theta$-angle at $\theta=\pi$ can have a mixed anomaly involving its time-reversal symmetry $\mathsf{T}$ and its 1-form symmetry $Z(\tilde{G})^{(1)}$.  Here we apply our \MRtext/ to constrain the possible long-distance behavior of this theory.  In particular, we show that for certain $G$ this theory cannot flow to a confined, gapped phase, with unbroken $\mathsf{T}$.

Before proceeding, we set conventions for the instanton numbers of various gauge groups.  For connected and simply-connected groups, the bundles on any four-manifold $M^4$ are labelled by a single characteristic class $\mathcal{I}$.  For these groups $\mathcal{I}$ is the integer-valued instanton number density
\begin{equation}\label{Idef}
	\mathcal{I}_G \equiv \frac{1}{4h^{\vee}} \mathrm{tr}\left(\frac{F}{2\pi}\wedge \frac{F}{2\pi}\right)~,
\end{equation}
where in the above $F$ denotes the curvature of the bundle, the trace is in the adjoint representation and $h^{\vee}$ is the dual Coxeter number. 

For non-simply-connected groups $G$, although the formula \eqref{Idef} still defines a bundle invariant, in general it is no longer integer-valued. Instead the fractional parts $\mathcal{I}_G^\text{frac} := (\mathcal{I}_G \mod 1) \in H^4(M^4,\mathbb{Q}/\Z)$ are fixed by the second Stiefel-Whitney class $w_2(G)$ as \cite{Witten:2000nv}\footnote{\label{spin4n}
When $G=\frac{\Spin(4N)}{\Z_2\times \Z_2}$, there are two independent obstructions valued in $H^2(M^4,\Z_2)$ to lift the bundle to a $\Spin$ bundle.  For simplicity we ignore this case but the interested reader can consult with \cite{Witten:2000nv, Cordova:2019uob} for further details.}
\begin{equation}
	\cI_G^\text{frac} = \frac{p_G}{q_{G}} \mathfrak{P}(w_2(G)) \mod 1~,
	\label{eq:Ifrac}
\end{equation}
where $w_2 \in H^2(M^4,\pi_1(G))$ is the characteristic class representing the obstruction to lifting the $G$-bundle to a bundle of its universal cover.  Here, $\mathfrak{P}$ is the Pontryagin square, which is the cohomology operation of type $H^2(-,\Z_k) \to H^4(-,\Z_{2k})$ when $k$ is even, and $\mathfrak{P}(x)=x \cup x$ for $x\in H^2(-,\Z_k)$ when $k$ is odd, and $p_G$ and $q_G$ are coprime integers depending on $G$ and specified in Table \cref{tab:pGqG} for classical groups \cite{Witten:2000nv, Cordova:2019uob}.
\begin{table}[t]
	\centering
	\begin{tabular}{c|c|c|c|c}
		$G$ & $\frac{\SU(N)}{\Z_{l}}$ & $\frac{\Spin(4N+2)}{\Z_4}$ & $\SO(N)$ & $\frac{\Sp(N)}{\Z_2}$\\[.5ex]
		\hline
		\rule{0pt}{2.5ex}$\frac{p_G}{q_G}$ & $\frac{N(N-1)}{2l^2}$ & $\frac{2N+1}{8}$ & $\frac12$ & $\frac{N}{4}$\\[.5ex]
		\hline
		\rule{0pt}{2.5ex}$q_G$ & $\frac{2l^2}{\gcd(2l^2,N(N-1))}$ & $8$ & $2$ & $\frac{4}{\gcd(N,4)}$
	\end{tabular}
	\caption{The coefficients in \eqref{eq:Ifrac} \cite{Witten:2000nv}. The $\SO(N)$ entry is valid for $N\ge 4$.}
	\label{tab:pGqG}
\end{table}

Consider the case with simply-connected $G$ (other than $\Spin(4N)$ for simplicity; see footnote \ref{spin4n}). The theta term in the action with $\theta=\pi$ is $\pi\mathrm{i}\int \cI_G$, which is invariant modulo $2\pi\mathrm{i}$ under the action of $\mathsf{T}$-symmetry, since $\int \cI_G$ is an integer.

However when the $Z(G)^{(1)}$ background $B$ is activated, the bundles summed over in the partition function are by definition bundles of the quotient group $G/Z(G)$ with fixed second Stiefel-Whitney class $w_2(G/Z(G)) = B$.  From \eqref{eq:Ifrac}, we can see that with a nontrivial background $B$ the term $\pi\mathrm{i}\int \cI_G$ is no longer invariant under $\mathsf{T}$.  Instead, the action of $\mathsf{T}$ generates a counterterm:\footnote{There are subtly different versions of pure Yang-Mills theory, see \cite{Wan:2018zql,Wan:2019oyr}, and the modification factor here can depend on the subtle choice when the 4-manifold is not spin. Here, however, this subtlety does not matter as we only consider the manifold $S^2\times S^2$, which is spin.}
\begin{equation}
\mathsf{T}\left(\exp(\pi\mathrm{i}\int \cI_G)\right)=\exp(\pi\mathrm{i}\int \cI_G+\tpi \frac{p_{\frac{G}{Z(G)}}}{q_{\frac{G}{Z(G)}}}\int \mathfrak{P}(B))~.
\end{equation}

To deduce whether this implies a non-trivial anomaly, we must check to see if this transformation can be eliminated by modifying the action by a suitable counterterm.  In this case, the term of interest is of the form $\frac{\tpi k}{|Z(G)|\gcd(|Z(G)|,2)}\int \mathfrak{P}(B).$   This is also not invariant under the $\mathsf{T}$ transformation and generates the variance $\frac{4\pi\mathrm{i} k}{|Z(G)|\gcd(|Z(G)|,2)}\int\mathfrak{P}(B)$. For the $\mathbb{Z}_2^{\mathsf{T}}\times Z(G)^{(1)}$ symmetry to be non-anomalous, the variance due to the counterterm should cancel the variance due to the theta term. For this to happen, we need an integer $k$ that solves:
\begin{equation}
	\frac{\tpi p_{\frac{G}{Z(G)}}}{q_{\frac{G}{Z(G)}}} = \frac{4\pi \mathrm{i} k}{|Z(G)| \gcd(|Z(G)|,2)} \mod \tpi~.
\end{equation}
This equation does not have a solution when $G = \SU(2N)$, $G = \Sp(2N+1)$, or $G = \Spin(4N+2)$. Therefore the pure Yang-Mills theory for these groups at $\theta = \pi$ has a $\Z_2^{\mathsf{T}}\times Z(G)^{(1)}$ mixed anomaly.

We now show that this mixed anomaly satisfies the hypothesis of our \MRtext/. We consider the four manifold $S^2\times S^2$ and take the background $B$ to be the sum of the fundamental classes of the each $S^2$ factor $B=[S^{2}\times \{pt\}]+[\{pt\}\times S^{2}]$. Then, the $\mathsf{T}$ transformation generates the anomalous sign
\begin{equation}
	e^{\frac{\pi\mathrm{i}}{2}\int_{S^2\times S^2} \mathfrak{P}(B)} = -1~.
\end{equation}
This means that, on the mapping torus of $S^2\times S^2$ twisted by the orientation reversing map of one of the two factors of $S^2$, the corresponding inflow action evaluates to be $-1$, and hence our \MRtext/ applies.\footnote{
The inflow action $\omega$ is studied for $G=\SU(2)$ in \cite{Wan:2018zql,Wan:2019oyr}, and it is $\frac14\beta\mathfrak{P}(B) = \frac14 (B\Sq^1B+\Sq^2\Sq^1B)$, where $\beta$ is the Bockstein operation associated to $\Z_2\to\Z_8\to\Z_4$, up to other terms irrelevant here. One can check that this reproduces the sign on the mapping torus considered here.}
Therefore, for the groups $G=\SU(2N), \Sp(2N+1), \Spin(4N+2)$, pure Yang-Mills theory at $\theta=\pi$ cannot have a confined gapped phase with unbroken $\mathsf{T}$.

\subsection{$4d$ Adjoint QCD}\label{sec:adjQCDdisc}

As a final example, we use our \MRtext/ to constrain the dynamics of $4d$ adjoint QCD.  We consider $\SU(N_{c})$ gauge theory with $N_{f}$ Majorana adjoint fermions.   The theory with gauge group $\SU(2)$ with $N_{f}=2$ was recently studied in \cite{Anber:2018tcj, Cordova:2018acb, Bi:2018xvr,  Wan:2018djl}  while other theories have been discussed in \cite{Shifman:2013yca, Poppitz:2019fnp} (See in particular \cite{Cordova:2018acb} for a detailed discussion of the anomalies. Also, see \cite{Unsal:2007vu,Unsal:2007jx} for earlier foundational results on adjoint QCD as well as \cite{Poppitz:2012sw,Poppitz:2012nz,Komargodski:2017smk,Shimizu:2017asf} for further results on adjoint QCD.)

Since there is only adjoint matter, this system has a global one-form symmetry $\mathbb{Z}_{N_{c}}^{(1)}.$  At the classical level, there is also a chiral $\uU(1)$ 0-form which is broken to $\mathbb{Z}_{2N_{f}N_{c}}$ by the ABJ-anomaly.  In particular, the $\mathbb{Z}_{2}$ subgroup of this discrete symmetry is fermion number $(-1)^{F}$.  The  $\mathbb{Z}_{2N_{f}N_{c}}$ 0-form symmetry has a mixed anomaly with the 1-form symmetry discussed above.  Let $x$ be the background field for the 0-form symmetry and $B$ the background field for the $1$-form symmetry.  Then, the inflow action is
\begin{equation}\label{anomadj}
	\tpi\omega = \frac{\tpi}{ \gcd(N_{c},2)N_{c}}\int_{M^5} x \cup \mathfrak{P}(B)~.
\end{equation}

This inflow action is nonzero on $M^5= S^1\times S^2 \times S^2$ with $x = a$ and $B= b_1 + b_2$ where $a,b_1,b_2$ are the fundamental classes of $S^1$ and the two copies of $S^2$:
\begin{equation}\label{anomevaladj}
\tpi\omega=\frac{\tpi}{ \gcd(N_{c},2)N_{c}}\int_{S^1\times S^2\times S^2} a\cup (b_1 + b_2)^2=\frac{4\pi i}{ \gcd(N_{c},2)N_{c}}~.
\end{equation}
Since the right-hand side above is non-trivial, we conclude that there is no TQFT that preserves both the chiral $\mathbb{Z}_{2N_{f}N_{c}}$ 0-form symmetry and is also confining, i.e. preserves the 1-form symmetry $\mathbb{Z}_{N_{c}}^{(1)}.$  This result extends the analysis of  \cite{Wan:2019oyr} which used a similar calculation to argue that there is no symmetry extension construction of such a TQFT.  In particular our result excludes certain exotic dualities proposed for these theories in \cite{Bi:2018xvr}. 

In fact, if we restrict our attention to confined, gapped phases we can use our \MRtext/ to deduce the required pattern of 0-form symmetry breaking.  Specifically, \eqref{anomevaladj} trivializes when $x$ above is taken to be a multiple of $N_{c}$ and therefore the minimal allowed breaking pattern is $\mathbb{Z}_{2 N_fN_{c}}\rightarrow \mathbb{Z}_{2N_f}$ thus leading to $N_{c}$ vacua.  This agrees with the symmetry breaking scenarios advocated in \cite{Shifman:2013yca, Anber:2018tcj, Cordova:2018acb, Poppitz:2019fnp}.

\section*{Acknowledgements}
We thank S.H. Shao and J. Wang for discussions.  C.C.\ was supported in part by DOE grant de-sc0009988. K.O.\ was supported in part by NSF Grant PHY-1606531 and the Paul Dirac fund.
\appendix

\section{Examples of Symmetry Preserving Gapped Phases}\label{apsymext}

In this appendix we review examples of symmetry-preserving TQFTs saturating non-trivial anomalies and see how they evade our  \MRtext/ .

\subsection{Discrete 0-from Symmetry Anomaly}

Let $G$ be a finite 0-form symmetry group with a bosonic anomaly $\omega \in H^{d+1}(BG,\RmZ)$, where $BG$ is the classifying space of $G$ bundles.  Given this data, \cite{Wang:2017loc} (see also \cite{Tachikawa:2017gyf}) constructed a finite group gauge theory that saturates the anomaly and preserves the symmetry in spacetime dimension $d\ge 3$.

Let us briefly review their construction.  Following \cite{Tachikawa:2017gyf}, we first note that given a finite group symmetry $G$ and its anomaly $\omega \in H^{d+1}(BG,\RmZ)$, it is possible to find an extension
\begin{equation}
	1 \to K \to H \to G \to 1~,
\end{equation}
where $K$ is abelian and the extension class $e \in H^2(BG,K)$ is such that there exists a class $b\in H^{d-1}(BG,\Hom(K,\RmZ))$ satisfying $\omega = b \cup e$.

We use this to produce a $K$ gauge theory saturating the anomaly $\omega$.  This gauge theory has global symmetry $G$ and hence couples to a background gauge field which we view as a (homotopy class of) map to the classifying space $g:M^d\to BG$.  The partition function is defined as 
\begin{equation}
	Z[M^d,g] \propto \sum_{\substack{\alpha \in C^1(M^d,K)\\\beta \in C^{d-2}(M^d,\hat{K})}}
	e^{\tpi\int_{M^d}(\beta \cup \delta \alpha + \beta \cup g^* e + g^* b \cup \alpha)}~,
	\label{eq:WWW}
\end{equation}
where $\hat{K} = \Hom(K,\RmZ)$ is the Pontryagin dual of $K$.

In the language of \cite{Benini:2018reh, WIP}, this $K$-gauge theory has intrinsic symmetry $K^{(1)} \times \hat{K}^{(d-2)}$ with the mixed anomaly $A\cup B$ where $A$ and $B$ are the backgrounds for the symmetry.  The partition function \eqref{eq:WWW} is constructed by making the theory couple to the $G$-background through $A= g^*e$ and $B = g^*b$.

As described in section \ref{secdiscex}, the intrinsic $K^{(1)}\times \hat{K}^{(d-2)}$ symmetry in the $K$-gauge theory is spontaneously broken. However, we must determine whether, when $G$ is coupled to the $K$ gauge theory as in \eqref{eq:WWW} the $G$ symmetry is spontaneously broken.  Clearly, when $d>2$ only higher-form symmetry is spontaneously broken in this gauge theory and hence the $0$-form symmetry $G$ is preserved.  On the other hand, if $d=2$, then $\hat{K}^{(d-2)}$ is an ordinary $0$-form symmetry.  Moreover $b\in H^1(BG,\hat{K})\simeq \Hom(G,\hat{K})$, so in this case the $G$ symmetry is broken down to the kernel of $b$.  In particular, this is consistent with the discussion in section \ref{sec:LMS}.

To see how this discussion is consistent with our \MRtext/.  Note that if $d\ge 3$, we are considering a mapping torus of the form $S^{1}\times_{f}(S^{k_{1}}\times S^{k_{2}})$ with at least one of $k_{i}>1$ and so we must show that all pure bosonic anomalies of zero form global symmetries are trivialized on manifolds of this form.  Indeed, as mentioned above, a $G$ gauge field can be viewed as a 
homotopy class of maps $g: M^{d} \to BG$.  The classifying space $BG$ has non-zero homotopy in dimension one, but $\pi_{\ell}(BG)$ is trivial if $\ell>1.$  In particular, this implies that when restricted to a sphere $S^{k}$ with $k>1$ the $G$ bundle trivializes, and thus the anomaly vanishes on the manifolds of interest.

\subsection{More on $4d$ Adjoint QCD}

Let us elaborate on the analysis of section \ref{sec:adjQCDdisc} which studied the anomalies of adjoint QCD.  We focus on the $\mathbb{Z}_{2N_{f}}$ subgroup of the 0-form symmetry, and denote by $z$ the background field for this $\mathbb{Z}_{2N_{f}}$.   The anomaly $\omega$ in \eqref{anomadj} is still non-trivial when $N_{c}$ is even and reduces to a mod 2 effect:
\begin{equation}\label{eq:xBB}
2 \pi i \omega = \frac{2 \pi i }{\gcd(N_{c},2)}\int z \cup B \cup B~.
\end{equation}
In particular $\omega$ depends only on $B$ modulo $2$ we can use $B\cup B=B\cup w_{2}~\mod 2,$ where $w_{2}$ is the second Stiefel-Whitney class of the spacetime manifold.  Since $S^{2}\times S^{2}$ is spin this gives another argument that the anomaly trivializes on this manifold.

Indeed, it is in fact straightforward to construct a TQFT with this anomaly where the symmetries are unbroken.  Specifically, consider the following $\Z_2^{(1)}$ gauge theory (this $\mathbb{Z}_{2}^{(1)}$ is a dynamical gauge group and should not be conflated with the background 1-form symmetry):
\begin{equation}
	Z[M^d,(x,B)] \propto \sum_{\substack{\alpha \in C^2(M^4,\Z_2)\\\beta \in C^{1}(M^4,\Z_2)}}
	e^{\frac{\tpi}{2}\int_{M^d}(\beta \cup \delta \alpha + \beta \cup x \cup B + w_2 \cup \alpha)}~,
	\label{eq:Z2bb}
\end{equation}
 where $w_2$ is the second Stiefel-Whitney class of the tangent bundle.  This saturates the anomaly and preserves $\Z_{2N_{f}}\times \Z_{N_{c}}^{(1)}$.  

The anomaly \eqref{eq:xBB} evaluates nontrivially on $S^1\times \mathbb{CP}^2$ with background $x = [S^1]$ and $B= w_2(T\mathbb{CP}^2)$.  Applying the logic of section \ref{sec:MappingTorus}, a QFT saturating this anomaly should have vanishing partition function on $\mathbb{CP}^2$ without any insertion of symmetry backgrounds:
\begin{equation}\label{cp20}
	Z[\mathbb{CP}^2] = 0~,
\end{equation}
which is indeed satisfied by the theory \eqref{eq:Z2bb}.  The vanishing \eqref{cp20}  is compatible with unitarity because $\mathbb{CP}^2$ does not have a reflection.\footnote{Any manifold with reflection should be null-bordant: see Lemma 4.1 of \cite{Yonekura:2018ufj}.}

\bibliography{biblio}{}
\bibliographystyle{utphys}

\end{document}